\documentclass{article}
\usepackage[utf8]{inputenc}
\usepackage{mathtools}
\usepackage{amsmath}
\usepackage{amsthm}
\usepackage{amsfonts}
\usepackage{amssymb}
\usepackage{comment}
\usepackage{appendix}
\usepackage{units}
\usepackage{comment}
\usepackage{authblk}
\usepackage{tikz}
\usepackage{lscape}
\usepackage{rotating}
\usepackage{hyperref}
\usepackage{adjustbox}

\usepackage[style=authoryear,sorting=nyt]{biblatex}
\usepackage{geometry}
\usepackage{csquotes}

\newtheorem{proposition}{Proposition}

\title{Gender-Segmented Labor Markets and Foreign Demand Shocks}
\author[1]{Carlos G\'oes\thanks{We thank Marc Muendler, Claudia Berg, and multiple participants of UCSD and World Bank seminars for their helpful comments. The very helpful reports from anonymous referees were essential to improve earlier versions of this paper. This paper served as background research to produce the World Bank's report ``Exports to Improve Labor Markets in the Middle East and North Africa.'' Financial support from the World Bank is gratefully acknowledged. The views expressed herein are those of the authors and do not necessarily reflect the views of the World Bank.}}
\affil[1]{UC San Diego}
\author[2]{Gladys Lopez-Acevedo}
\affil[2]{World Bank}
\author[3]{Raymond Robertson}
\affil[3]{Texas A\&M University}

\date{\today}

\begin{document}

\maketitle

\begin{abstract}
Gender segmentation in labor markets shapes the local effects of international trade. We develop a theory that combines exports with gender-segmented labor markets and show that, in this framework, foreign demand shocks may either increase or decrease the female-to-male employment ratio. If a foreign demand shock happens in a female-intensive (male-intensive) sector, the model predicts that the female-to-male employment ratio should increase (decrease). We then use plausibly exogenous variation in the exposure of Tunisian local labor markets to foreign demand shocks and show that the empirical results are consistent with the theoretical prediction. In Tunisia, a developing country with a high degree of gender segmentation in labor markets, foreign-demand shocks have been relatively larger in male-intensive sectors. This induced a decrease in the female-to-male employment ratio, with households likely substituting female for male labor supply.\end{abstract}



\newpage

\section{Introduction}

Gender inequality in labor markets is ubiquitous. Even in countries that rank high in gender equality indicators, female workers tend to earn less than males with the same observable characteristics\footnote{For a review of gender imbalances in labor markets of developed countries, see the review in \textcite{oecd_closing_2012}.}. Such persistent differences in economic outcomes between men and women, with otherwise similar skills and education, can reduce economic growth \parencite{duflo_women_2012} and contribute to social instability \parencite{hudson_hillary_2015}. Biases, social norms, and/or individual preferences represent socio-economic forces that concentrate men and women in different industries and occupations \parencite{papageorgiou_economic_2018}, resulting in frictions that prevent workers from optimally moving across sectors. If large enough, such frictions can lead to substantial levels of market segmentation – a situation in which male and female workers perform different sets of pre-determined tasks, rather than sorting into tasks at which they are more productive. 

International trade affects men and women differently. Trade shocks induced by changes in policy or foreign demand alter domestic relative prices, both in product and factor markets, and impact both employment and wages. If the domestic market has a high degree of gender segmentation, it would not be surprising that trade shocks impact males and females in different ways. Since economic differences between men and women are pervasive, understanding how those differences interact with trade shocks to produce potentially unequal labor-market outcomes for men and women is relevant both as a scientific question and for its policy implications. 

In this paper, we make two main contributions to the literature: one theoretical and another empirical. First, we present a general equilibrium model that combines international trade (and exports in particular) and gender-segmented labor markets. We presume that social norms, technological-differences, and institutions might condition the kinds of tasks that are acceptable for each gender such that males and females are seen as different kinds of workers and different industries have different gender-specific labor demand patterns. Production is divided between a female-intensive sector and a male-intensive sector, and households optimize their combined consumption and leisure by supplying different quantities of female and male labor, given wages. In this framework, a foreign demand shock can either increase or decrease the equilibrium female-to-male employment ratio. The response depends crucially on (i) the sectoral composition of the demand shock; and (ii) the relative market shares of each industry in the destination market.

We then present empirical evidence consistent with this theoretical mechanism. We construct a new dataset that combines labor market data and trade data from Tunisia, a developing country with a high degree of gender segmentation in the labor market and where foreign demand shocks were concentrated in male-intensive industries. We then use plausibly exogenous variation in export exposure of different Tunisia regions to show that the direction of the effect of exports on labor markets is the same as the one predicted by the theoretical model. 

Methodologically, our empirical approach is similar to \textcite{autor_china_2013}, \textcite{malgouyres_impact_2017}, and 
\textcite{benguria_decomposing_2023}, who focus on the effects of imports on local labor markets. However, our focus is on exports, which is a channel that is under-emphasized in the empirical literature. To implement our empirical strategy, we instrument export growth with a weighted average of foreign demand shocks, relying on the fact that Tunisia is a small open economy (similar to \cite{aghion_heterogeneous_2022}). In combining these two strategies, we are able to use local-labor markets differential exposure to \textit{exports}, instrumented by foreign demand shocks, to test our hypothesis.

Tunisia is a good case study because it has a high level of gender segmentation in labor markets but females are an important part of the labor force, participating at higher rates than other Middle East and Northern Africa (MENA) countries. During the sample period, the female-to-male employment ratio decreases in response to a foreign demand shock. This effect was concentrated particularly among married female workers, but there is not differential response between married or single male workers, which suggests that households substitute male for female labor supply when male-intensive industries expand.

\paragraph{Relationships with the Literature.} The literature does not come to clear conclusions for how trade shocks should affect relative economic outcomes for men and women. Differences in theoretic approaches and empirical settings generate a wide range of results that sometimes contradict one another. The theoretic approaches economists have applied fall into four broad categories.

The first, inspired by Becker’s theory of ``taste-based discrimination'' \parencite{becker_economics_1971}, posits that industries inherently prefer male workers.  The preference for men generates a wage differential in favor of men that cannot be easily sustained in a competitive market. Therefore, the male-female wage and employment gaps are higher in concentrated industries.  Trade liberalization increases competition and reduces the ability of concentrated industries to discriminate.  As a result, trade is associated with falling wage and employment gaps.  Studies that apply this approach, such as \textcite{black_importing_2004}, \textcite{berik_international_2004}, \textcite{menon_international_2009}, and \textcite{paz_effects_2021}, find empirical support for this approach.

The second approach, based on a household labor supply model (e.g. \textcite{stephens_worker_2002}; \textcite{attanasio_female_2005}, \textcite{attanasio_explaining_2008}, and \textcite{donni_collective_2018}), posits that when one spouse loses a job, the other is more likely to enter the labor market.  If, for example, imports disproportionately induce job losses for men, women are more likely to enter the labor market to maintain household income \textemdash the ``added worker effect.''  At the same time, a loss in earning power for one spouse makes it more likely that the affected spouse will drop out of the formal labor market and concentrate time on household production while the other would likely enter or increase hours in the formal labor market.

The third approach suggests that men and women differ in their degree of complementarity with technology or capital. Inspired by \textcite{goldin_understanding_1990}, several studies such as \textcite{juhn_men_2014}, \textcite{ben_yahmed_gender_2020}, and \textcite{saure_international_2014} introduces models in which women and capital or technology are complements. \textcite{juhn_men_2014} and \textcite{ben_yahmed_gender_2020} argue that falling gender wage or employment gaps in Mexico were due to women’s complementarity with technology or capital. In the U.S., \textcite{saure_international_2014} argue that the relative contraction of the male-intensive sector due to trade liberalization with Mexico induced men to move into the female-intensive sectors, lowering the marginal product of, and therefore demand for, women.  Earlier, Kongar \parencite*{kongar_importing_2006} reached a similar conclusion by showing that imports reduced the demand for low-skilled workers that disproportionately included women.  \textcite{kongar_importing_2006}, however, found that the wage gap fell as low-wage women lost jobs, while \textcite{saure_international_2014} suggest that the economic gap between men and women grew. \textcite{autor_when_2019} show the (cross-sectional) gender wage gap decreased in the United States due to Chinese import competition, as manufacturing was more male-intensive, and this result holds throughout the income distribution. 

The fourth approach assumes that, possibly due to social norms or legal restrictions, otherwise identical men and women are not perfect substitutes in production. This assumption has been applied widely in a number of studies including, but not necessarily limited to, \textcite{grant_labor_1980}, \textcite{galor_gender_1993}, \textcite{de_giorgi_gender_2013}, \textcite{olivetti_economic_2017}, \textcite{saure_international_2014} and \textcite{do_comparative_2016}. In this approach, the elasticity of substitution of otherwise identical men and women, regardless of factor complementarity, is finite and small.  The aforementioned social institutions lead to occupational and industrial segregation that determines how trade affects men and women differently \parencite{brussevich_does_2018}.  

Our theoretical model shares similarities with all four different strands of the literature. We posit that due to biases, technological complementarities, social norms or preferences labor markets are gender segmented, which induce male and female workers to be perceived as different kinds of workers in the labor force.\footnote{We do not endogenize or model why these differences emerge, but rather take them as given, since our focus is to understand the economic consequences of trade shocks \textit{given gender segmented labor markets}.}  Male and female workers are also imperfect substitutes in the labor market and are employed as factors of production in both industries. Finally, households optimize the relative supply of male or female labor supply depending on relative wages and their specific household budget, which can account for the (negative income) added worker effect.

More closely related to our theoretical approach is \textcite{do_comparative_2016}, who put forth a theory stating that countries that have a comparative advantage on female intensive sectors are characterized by lower fertility. However, our frameworks differ in some key aspects. First, their focus is on the household's marginal choice regarding female labor versus fertility, assuming that males always supply labor inelastically. Conversely, our model makes predictions regarding the household's choice of male versus female labor supply and abstracts away from fertility decisions. Second, while their model predicts a stark specialization of genders across sectors, we accommodate a milder assumption that sectors differ in their gender-intensity but both males and female workers are inputs in all sectors of the economy. Finally, our focus is on how gender-segmented labor markets respond to foreign demand shocks, while their focus relates to Ricardian comparative advantage in female- vs male-intensive sectors and fertility decisions.

Empirical results also vary across studies. For example, the World Bank’s (2020) Women and Trade report shows that commodities sectors are generally male-intensive relative to other traded goods and services, and that countries that export commodities tend to have higher gender wage gaps throughout the economy.  Shifting exports away from commodities, therefore, should increase the relative demand for women and improve female labor market outcomes. \textcite{oostendorp_globalization_2009} found that rising trade is associated with falling occupational (across industry) gender wage gaps in a set of countries.  More specifically, in Brazil, \textcite{chauvin_gender-segmented_2018} finds that the effects of local demand shocks depend on whether they favor women due to industrial segmentation. In Honduras, De \textcite{de_hoyos_exports_2012} found that rising exports of Honduras’s maquila sector, of which about 70 percent of workers are female, reduced poverty by increasing women’s wages and employment.  

In Tunisia, during the sample period, male-intensive sectors faced relatively larger foreign demand shocks. As a consequence, consistent with our theoretical model, the female-to-male employment ratio decreases in response to a foreign demand shock that is skewed toward the male-intensive sector. This effect is concentrated particularly among married female workers, which suggests that households substitute male for female labor supply when male-intensive industries expand. Therefore, our results are consistent with previous estimates that suggest that in \textit{relative terms} export shocks increase relative female employment. One qualification that we add to these studies is that not all exports are equal. The nature of the foreign demand shocks, which sectors in the economy are expanding, and their gender intensity matter. One common caveat across all of these studies (that we also share) is that all of these relative estimates do not capture level effects that jointly affect both comparison groups. We discuss this in further detail in the empirical and caveats sections of the paper.

Mobility between occupations, industries, and particularly across regions appears to be limited. The occupational and industrial segregation implied by this approach, however, also implies that mobility between occupations and industries is limited and these frictions can amplify the effects of trade liberalization \parencite{dix-carneiro_trade_2017}.  We find little evidence of migration across regions in Tunisia as a response to foreign-demand shocks.  This is important because it implies that adjustment to a negative shock to a female-intensive industry will reduce female labor force participation. After all, women are not able to move out of the contracting industry and into the expanding industry. Studies that focus on trade, reallocation, and female labor force participation find results consistent with this prediction. \textcite{mansour_import_2020} and \textcite{heckl_import_2023}, for example, find that rising Chinese imports (significantly, of apparel) into Peru and Mexico, respectively, reduced demand for women’s labor.  The contraction in female employment was roughly twice that for men in Peru, and, in both countries, female labor-force participation fell. On the other hand, imports from China reduced the relative demand for less-educated males in  the United States and was followed by an increase in female labor force participation as the economy shifted towards female-intensive sectors \parencite{besedes_trade_2021}.  \textcite{kis-katos_globalization_2018} analyze tariff liberalization in Indonesia and found that female-intensive sectors expanded after a tariff liberalization, female work participation increased, and the time women spent on domestic duties decreased. \textcite{gaddis_gendered_2017} use data from Brazil and a difference-in-differences strategy to show that trade liberalization reduced the relative (cross-sectional) gender employment gap. We find little evidence of migration across regions in Tunisia as a response to foreign demand shocks.

It is important to point out that nearly all of the previous gender-trade studies focuses on imports.  Few studies focus on exports.  Exports are important because China may have displaced Tunisia’s exports in the European Union, especially in apparel, highlighting another important dimension of the ``China Syndrome''.  \textcite{robertson_labour_2020} show that competition with China in the U.S. market, especially in apparel, displaced women in Mexico.  Other studies focus on Brazil. \textcite{costa_winners_2016} find that exports have important positive labor-market effects that are significantly correlated with the composition of exports. \textcite{paz_effects_2021} find no relationship between industry-level export shares and female employment or the wage gap. In contrast,  \textcite{connolly_effects_2022} compares import and export effects and finds that overall trade between Brazil and China generally favored women over men in Brazil, especially in terms of employment.  \textcite{robertson_mending_2022} also present a model of heterogenous firms in which firms choose between male and female labor as inputs and respond to exogenous price shocks in Bangladesh’s export market.  We therefore expect that our paper will contribute to an emerging picture of heterogeneous effects of trade that are at least partially explained by differences in the composition of trade and degree of labor market segmentation.

One final connection to the literature that our paper makes regards gender and labor markets in the Middle East and North African (MENA) region. Differences in economic outcomes between men and women are especially stark in MENA.  Like many developing countries, MENA countries have significantly liberalized trade since 2000 with the hopes of improving economic outcomes. A number of different relevant studies have focused on MENA and Tunisia. \textcite{hyland_gendered_2020} use novel data from a World Bank survey and document that gender inequities regarding ``pay and treatment of parenthood'' are relatively higher in MENA compared to other regions in the world, and those have not improved substantially over the last decades. \textcite{miller_integration_2022} match data from the World Bank Enterprise Surveys and theory to show that integration costs across genders are higher in MENA firms. \textcite{friedrich_womens_2021} use labor force data from Egypt, Jordan, and Tunisia to highlight the existence of a paradox: while educational attainment improved and fertility rates declined in MENA, female financial autonomy and individual freedom (including work outside the household) did not improve. \textcite{alaref_medium-term_2020} focus on educational outcomes, showing that a randomized experiment that increases female exposure to entrepreneurial classes only had short-lived results in Tunisia.  \textcite{kaasolu_female_2019} studied the Jordanian case in depth and showed that female labor force participation varies substantially with educational attainment, with social norms constraining less educated women more significantly.

We add to this literature by documenting how international trade and labor markets interplay in an institutional setting of highly segmented labor markets. We first show that export growth can be skewed towards male- or female-intensive industries, and, in the case of Tunisia, male-intensive industries expanded. Through the lens of a structural model, we then argue that the relevance of the destination markets and the gender-intensity of an industry will shape the nature of the labor market response. While Tunisia was our case-study, this topic is clearly of broader interest to MENA countries, which have institutional settings more closely related to Tunisia.

\section{Theory}

Significant labor-market gender segmentation suggests that there could be frictions that discourage women from moving across industries. Formal or informal institutions – such as social norms – can make males and females be perceived as different kinds of workers that contribute in different proportions as factors of production in the labor market. One way to rationalize this framework is to assume that social norms dictate that some tasks are "female" tasks (e.g., sewing for garments) while other tasks are "male" tasks (e.g., operating an oil rig). If industries use tasks in different proportions, they will be either female- or male-intensive industries.

Gender segmentation in the labor market implies that households now face a decision regarding whether to substitute male for female labor supply after facing a shock. For instance, if a male worker is laid off in a household, the female worker might want to increase her labor supply outside of the household in order to supplement the household’s budget. Conversely, if wages in male-intensive industries increase, females might be induced to reduce their work outside of the household, since the relative opportunity cost of housework becomes lower.

Since many developing countries, including Tunisia, are characterized by gender-segmentation in labor markets, we devise a model in which sectors use male and female labor as inputs but with different factor intensities. The main result is a proposition that predicts that the response of the female-to-male employment ratio in the export market depends on the male-intensity of the demanded good in the importing country and the relevance of that market to the export market. Depending on the parametrization of the model, some differences arise regarding the relative expansion and contraction of either sector and how the model contrasts with canonical trade models. We discuss these differences at the end of this section.

\paragraph{Demand} We consider a world economy with several regions $d \in \mathcal{K}$ that may or may not be part of the same country. In each region, there is a representative household that maximizes utility by choosing optimal consumption as well as female and male labor supply. For simplicity assume that migration costs across regions are prohibitively high\footnote{In the empirical part of the paper we assess the relevance of within country migration for identifying empirical results.}.

\begin{eqnarray*}
    &\max_{ \{C_{d,m},C_{d,m},L_{d,m},L_{d,f} \}}& C_{d,m}^{\alpha_d} C_{d,f}^{1-\alpha_d} - \frac{L_{d,m}^{1+\eta_d}}{1+\eta_d} - \nu_d\frac{L_{d,f}^{1+\eta_d}}{1+\eta_d}  \\
    &s.t.& P_{d,m}C_{d,m}+P_{d,f}C_{d,f}\le L_{d,m}w_{d,m}+L_{d,f}w_{d,f}+ P_d e_d=Y_d
\end{eqnarray*}

\noindent where $C_{d,m},C_{d,f}$ are quantities demanded for goods $m$ and $f$, respectively, in country $d$; $P_{d,m},P_{d,f}$ are the prices of those two goods; $L_{d,m}$ is female labor supply; $L_{d,f}$ is male labor supply; and $e_d$ is some endowment, which can be interpreted as an income shifter. Given this structure, we can define the ideal price index $P_d\ =\ A\cdot P_{d,m}^{\alpha_d}P_{d,f}^{1-\alpha_d}$, where $A = \left(\alpha_d\right)^{-\alpha_d}\left({1-\alpha}_d\right)^{-(1-\alpha_d)}$. 

As it is clear from the formulation above, by assumption, males and females are \textit{different kinds of workers}. Here, we summarize this heterogeneity in a synthetic way, by proposing an economy with two sectors: a male-intensive one that produces good \textit{m} and a female-intensive one that produces good \textit{f}\footnote{One can endogenize the product shares $\alpha_d$ by assuming the existence of a continuum of industries $i$ in which goods are produced with a linear technology (i.e., female and male workers are substitutes) and the relative productivity of genders is monotonic in $i$, such that genders specialize in a range of the industries space. We chose this simpler framework because the mechanism between foreign demand shocks and relative wages for different genders would be very similar. Furthermore, our focus is not to understand why labor markets are segmented, but rather to, given labor market segmentation, understand the impact of foreign demand shocks on domestic labor markets.}. Social norms may also influence household preferences for supplying female versus male labor outside of the household, as summarized by the parameter $\nu_d$.

\paragraph{Production} In each of those sectors $i\ \in\{m,f\}$, there is a domestic competitive firm that aggregates country-specific differentiated varieties $c_{od,i}$ into a composite good: 
\begin{equation*}
    C_{d,i} \equiv \left(\sum_{o\in\mathcal{K}}\left(c_{od,i}\right)^\frac{\sigma-1}{\sigma}\right)^\frac{\sigma}{\sigma-1}
\end{equation*}

\noindent where $\sigma > 0$ is the elasticity of substitution across varieties. This technology implies that the price $P_{d,i}$ of each composite good satisfies:

\begin{equation*}
    P_{d,i}\equiv\left(\sum_{o\in\mathcal{K}}\left(p_{od,i}\right)^{1-\sigma}\right)^\frac{1}{1-\sigma}
\end{equation*}

\noindent where $p_{od,i}$ is landed price of each variety. Demand for each differentiated variety satisfies:

\begin{equation*}
    c_{od,i}=\left(\frac{p_{od,i}}{P_{d,i}}\right)^{-\sigma}C_{d,i}
\end{equation*}

Firms in sector $i$ use female labor or male labor to produce their variety according to a constant-returns to scale technology that combines female and male labor in different intensities:

\begin{equation*}
    c_{od,i}=z_{o,i} \left( \ell_{m,od}^i \right)^{\beta_{o,i}} \left( \ell_{f,od}^i \right)^{1-\beta_{o,i}}
\end{equation*}

\noindent Since $m$ is the male-intensive sector, we assume $\beta_{o,m} > \beta_{o,f}$ for all $o \in \mathcal{K}$. Here $z_{o,i}$ is total factor productivity in sector $i$ of origin country $o$; $\ell_{j,od}^i$ is labor demand for worker of type $j\ \in\{m,f\}$.

\paragraph{Trade equilibrium} We further assume that there is free entry and perfect competition in both sectors, such that prices equal their marginal costs.  However, in this world economy, trade is not costless. If households in region $d$ want to consume one unit of a given variety from country $o$, they must source and pay for $\tau_{od}\geq1$ units – i.e., consumers face iceberg trade costs. We make the standard assumption that $\tau_{oo}=1$ (self-trade is costless) and $\tau_{od}\le\tau_{oz}\tau_{zd}$ (trade costs satisfy the triangle inequality). Therefore, landed prices in sector $i$ satisfy:

\begin{equation*}
    p_{od,m}=\frac{\tau_{od}(w_{o,m})^{\beta_{o,m}} (w_{o,f})^{1-\beta_{o,m}}}{z_{o,m}}, \qquad p_{od,f}=\frac{\tau_{od}(w_{o,m})^{\beta_{o,f}} (w_{o,f})^{1-\beta_{o,f}}}{z_{o,f}}
\end{equation*}

Note that we can rewrite the expenditure of consumers in $d$ on goods coming country $o$ in sectors $m,f$ as:
\begin{eqnarray*}
X_{od,m} &\equiv& p_{od,m}c_{od,m} = \left(\frac{p_{od,m}}{P_{d,m}}\right)^{1-\sigma}P_{d,m}C_{d,m}=\left(\frac{p_{od,m}}{P_{d,m}}\right)^{1-\sigma}\alpha_dY_d \\
X_{od,f} &\equiv& p_{od,f}c_{od,f}=\left(\frac{p_{od,f}}{P_{d,f}}\right)^{1-\sigma}P_{d,f}C_{d,f}=\left(\frac{p_{od,f}}{P_{d,f}}\right)^{1-\sigma}{(1-\alpha}_d)Y_d
\end{eqnarray*}

\noindent where the last equality comes from the Cobb-Douglas structure of preferences. We can also synthetically express expenditure in goods of sector $i \in \{m,f\}$ coming from $o$ as a share of total expenditure of country $d$ in that sector:
\begin{eqnarray*}
\pi_{od,m} &\equiv& \frac{X_{od,m}}{X_{d,m}}=\frac{\left(\tau_{od}(w_{o,m})^{\beta_{o,m}} (w_{o,f})^{1-\beta_{o,m}}\right)^{1-\sigma}\left(z_{o,m}\right)^{\sigma-1}}{\sum_{k\in\mathcal{K}}{\left(\tau_{kd} (w_{k,m})^{\beta_{k,m}} (w_{k,f})^{1-\beta_{k,m}} \right)^{1-\sigma}\left(z_{k,m}\right)^{\sigma-1}}} \\
\pi_{od,f} &\equiv& \frac{X_{od,f}}{X_{d,f}}=\frac{\left(\tau_{od} (w_{o,m})^{\beta_{o,f}} (w_{o,f})^{1-\beta_{o,f}} \right)^{1-\sigma}\left(z_{o,f}\right)^{\sigma-1}}{\sum_{k\in\mathcal{K}}{\left(\tau_{kd} (w_{k,m})^{\beta_{k,f}} (w_{k,f})^{1-\beta_{k,f}} \right)^{1-\sigma}\left(z_{k,f}\right)^{\sigma-1}}}
\end{eqnarray*}

\noindent where, by assumption, total expenditure in goods of either sector is $X_{d,m}=\alpha_dY_d, X_{d,f}=(1-\alpha_d)Y_d$.  Total labor income in each sector in origin country $o$ satisfies:
\begin{eqnarray*}
L_{o,f}w_{o,f} &=& (1-\beta_{o,f}) \sum_{d\in\mathcal{K}}\pi_{od,f}\cdot(1-\alpha_d)Y_d + (1-\beta_{o,m}) \sum_{d\in\mathcal{K}}\pi_{od,m}\cdot \alpha_d Y_d  \\
L_{o,m}w_{o,m} &=& \beta_{o,f} \sum_{d\in\mathcal{K}}\pi_{od,f}\cdot(1-\alpha_d)Y_d + \beta_{o,m} \sum_{d\in\mathcal{K}}\pi_{od,m}\cdot \alpha_d Y_d \\
Y_d &=& L_{d,m}w_{d,m}+L_{d,f}w_{d,f}+P_de_d \\
\end{eqnarray*}
\noindent which, given labor allocations, represents a system of 3N nonlinear equations with 3N endogenous variables $(w_{1,f},\ \cdots,\ w_{N,f},w_{1,m},\ \cdots,\ w_{N,m},Y_1,\ \cdots,\ Y_N)$ and solves for the trade equilibrium of this world economy up to the choice of a numeraire. 

\paragraph{Labor Market} To complete the characterization of this world economy, one needs to specify 2N labor market clearing equations that, combined with the 3N equations from the trade equilibrium, solve for world equilibrium in this economy. From the optimal decisions of the household, we can show that the ratio of female-to-male labor supply is increasing in the wage ratio with elasticity $1/\eta_o$:
\begin{equation*}
    L^s\equiv\frac{L_{o,f}}{L_{o,m}}=\left(\frac{1}{\nu_o}\frac{w_{o,f}}{w_{o,m}}\right)^\frac{1}{\eta_o}
\end{equation*}

Intuitively, since households are choosing optimally which bundle of female and male labor to supply, the ratio of female-to-male labor supply is an increasing function of relative wages. The taste parameter $\nu_o$ rotates the relative labor supply curve up or down around the origin. Conversely, relative labor demand is decreasing in the relative wage:

\begin{equation*}
    L^d\equiv\frac{L_{o,f}}{L_{o,m}}=\frac{ (1-\beta_{o,f}) \sum_{d\in\mathcal{K}}\pi_{od,f}\cdot(1-\alpha_d)Y_d + (1-\beta_{o,m}) \sum_{d\in\mathcal{K}}\pi_{od,m}\cdot \alpha_d Y_d  }{\beta_{o,f} \sum_{d\in\mathcal{K}}\pi_{od,f}\cdot(1-\alpha_d)Y_d + \beta_{o,m} \sum_{d\in\mathcal{K}}\pi_{od,m}\cdot \alpha_d Y_d }\cdot\left(\frac{w_{o,f}}{w_{o,m}}\right)^{-1}
\end{equation*}

\noindent which is decreasing in the relative wage. Here, a foreign demand shock can shift the labor supply curve either up or down. The direction of the shift depends on the sectoral composition of the shock and trade shares in a given market. We summarize the result in the proposition below.

\begin{proposition}[Effects of foreign demand shocks on a gender segmented labor market]\label{prop: 1} Let the world economy be as described above and define $\Xi_o\equiv\beta_{o,f} \sum_{d\in\mathcal{K}}\pi_{od,f}\cdot(1-\alpha_d)Y_d + \beta_{o,m} \sum_{d\in\mathcal{K}}\pi_{od,m}\cdot \alpha_d Y_d$. Suppose that factor intensities satisfy $0 \le \beta_{o,f} < \frac{\Xi_{o}}{1+\Xi_o} <  \beta_{o,m} \le 1$. Then a foreign demand shock, defined as an increase $e_d$ for some arbitrary country $d \neq o$, will decrease the female-to-male employment ratio $L_{o,f} / L_{o,m}$ if and only if (a) consumption in the destination market is sufficiently male-intensive ($\alpha_d$ is sufficiently large); and (b) the destination market in which the demand shock originates is a sufficiently important destination to the origin market ($\pi_{od,m}$ is sufficiently large). Otherwise, the female-to-male employment ratio $L_{o,f} / L_{o,m}$ will increase. 
\end{proposition}

\begin{proof}
    Appendix
\end{proof}

\paragraph{Discussion}

Proposition \eqref{prop: 1} shows that the way the female-to-male labor ratio responds to a foreign demand shock depends crucially on (a) the sectoral composition of the demand shock; and (b) the relative market shares of each industry of region $o$ in the destination market $d$. The former determines whether the demand shock from region $d$ translates into higher demand for goods whose production is female or male intensive. The latter controls how intensely the demand increase in $d$ transmits to origin labor markets $o$.

Suppose that there is a large shift in \textit{relative} demand away from female-intensive industries, as it was indeed the case in Tunisia during the time period that we analyze. As a stylized fact, the apparel sector, which tends to be female-intensive all over the world, and also in Tunisia, decreased its share in total Tunisia exports from 26\% of total exports in 2006 to 16\% of total exports in 2016. The same happened with the electrical machinery, the third most female-intensive sector in the Tunisian economy, which decreased its exports share from 20\% to 12\%. We discuss these data in more detail in the next section.

\begin{figure}[htp!]
    \centering
    \includegraphics[scale=0.5]{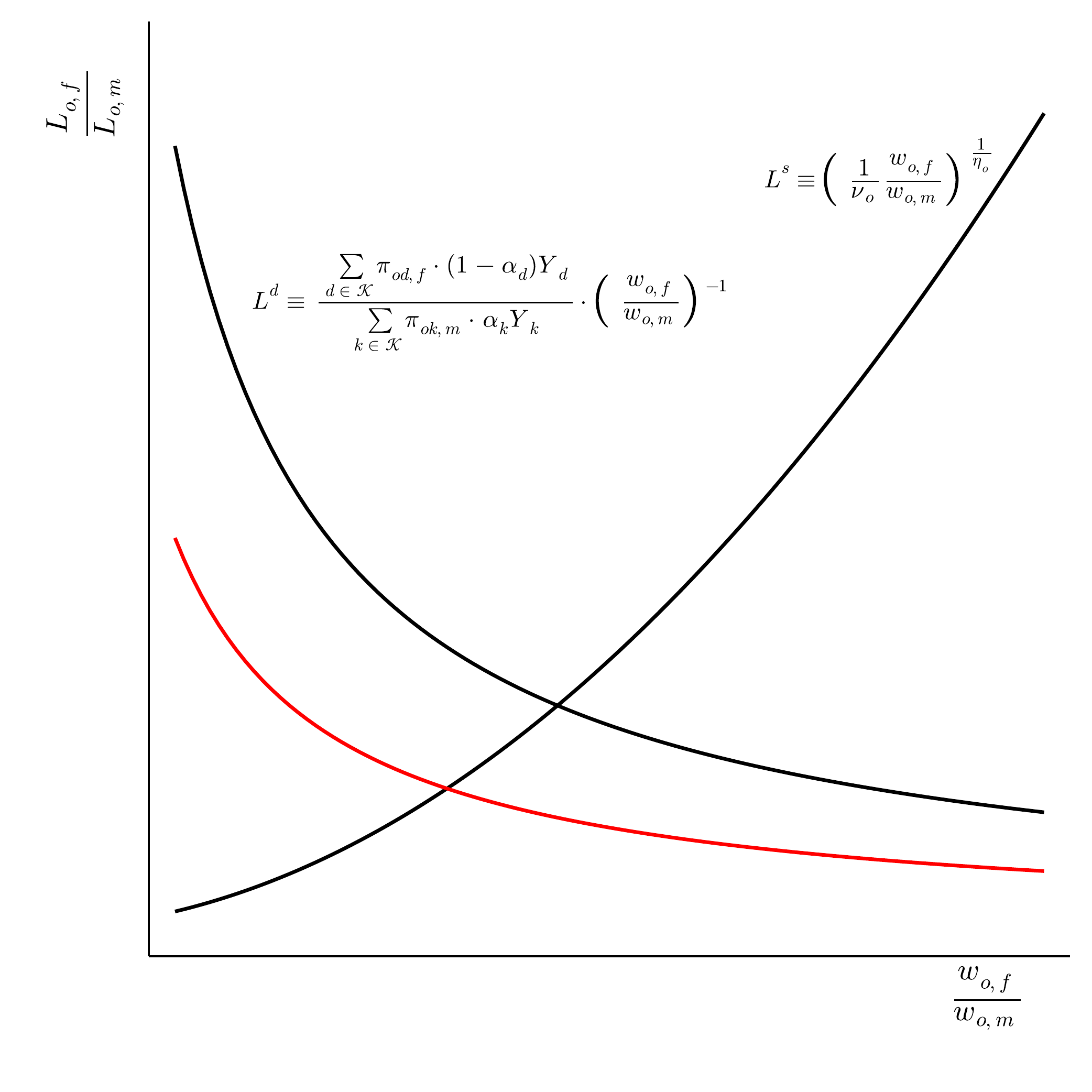}
    \caption{\textbf{Labor Market Equilibrium}. This chart illustrates the result of Proposition \eqref{prop: 1} in the special case that sectors are fully specialized \textemdash $\beta_{o,m} = 1$ and $\beta_{o,f} = 0$. A sufficiently large increase in the foreign relative demand for male-intensive goods will increase the labor demand for males relative to females, shifting the relative labor demand curve down. The resulting equilibrium is a lower female-to-male employment ratio.}
    \label{fig:LsLd}
\end{figure}

Intuitively, a sufficiently large decrease in foreign demand for female-intensive goods will decrease the labor demand for female workers relative to males. All else equal, this will decrease the relative wages of female workers. In equilibrium, households respond by increasing the male labor supply or decreasing the female labor supply up to the point at which the labor market clears. This result is illustrated in Figure \ref{fig:LsLd}.

The term $\Xi_o = L_{o,m}w_{o,m}$ corresponds to male household income and pins down the minimum amount of gender segmentation necessary for the result in the proposition to hold. In intuitive terms, what this condition is saying is that if male share of household income is close to zero or close to one, sectors must be sufficiently segmented for the result to hold.

Depending on the parametrization, some of the theoretical predictions of the baseline model are exactly the opposite of those from a standard Heckscher-Ohlin (HO) model with male and female labor as the two common factors. In that model, after a positive shock to the male-intensive sector, the aggregate female-to-male labor ratio is \textit{unchanged}, since factor supply is assumed to be exogenous and the HO model assumes full employment. Both sectors, however, would \textit{increase} their female-to-male labor use ratios: male workers are in high demand, but in general equilibrium both industries substitute away from male labor because their relative wages also increased just enough to maintain full employment.

By contrast, consider the model will full gender segmentation, i.e. in which each sector only hires one gender and $\beta_{o,m}=1$ and $\beta_{o,f}=0$. With such a kind of stark gender segmentation in labor markets, after a positive foreign demand shock to the male-intensive sector, the aggregate female-to-male \textit{decreases} and sector factor use ratios are \textit{unchanged}, since by assumption each sector only uses one type of labor: exactly the opposite predictions of the Heckscher-Ohlin model. 

Now suppose that both sectors use male and female labor as inputs, but do so with differing intensities, i.e., $\beta_{o,m}, \beta_{o,f} \in (0,1)$. Both sectors would \textit{increase} their female-to-male labor use ratios, like in the Heckscher-Ohlin model.  Since total labor factor supply is fixed, however, the relative size of the male-intensive sector would increase in a way that would more than compensate for changes in factor use ratios in a similar way to what happens in the Heckscher-Ohlin model.

Therefore, in equilibrium, if a shock induces a \textbf{decrease} in the female-to-male employment ratio and a decrease in the relative wages, then (a) both sectors \textbf{increase} their female-to-male labor ratio use; and (b) the male-intensive sector must expand in a way that more than compensates the increase in factor intensities. 

To see that, first note that, within each sector, the factor use ratio is equal to:

\begin{eqnarray*}
    \frac{L_{o,f}^i}{L_{o,m}^i} = \frac{\beta_{o,i}}{1-\beta_{o,i}} \left( \frac{w_{o,f}}{w_{o,m}} \right)^{-1}, \qquad \text{for } i \in \{m,f\}
\end{eqnarray*}

\noindent which shows that, in equilibrium, if $\frac{w_{o,f}}{w_{o,m}}$ decreases, $\frac{L_{o,f}^i}{L_{o,m}^i}$ must increase. Furthermore, the aggregate labor ratio can be written as:

\begin{eqnarray*}
    \frac{L_{o,f}}{L_{o,m}} = \mu_{o,m} \frac{L_{o,f}^m}{L_{o,m}^m}  +  (1-\mu_{o,m}) \frac{L^f_{o,f}}{L^f_{o,m}}, \qquad \mu_{o,m} \equiv \frac{L_{o,f}^m}{L_{o,m}}, \quad L_{o,m} = L_{o,f}^m + L_{o,f}^f
\end{eqnarray*}

By Proposition \eqref{prop: 1} $\frac{L_{o,f}}{L_{o,m}}$ decreases after a sufficiently male-intensive foreign demand shock. By the result previously stated, $\frac{L_{o,f}^i}{L_{o,m}^i}$ increases in both sectors after the same shock. To reconcile these facts, since $\frac{L_{o,f}^m}{L_{o,m}^m} > \frac{L^f_{o,f}}{L^f_{o,m}}$, the male intensive sector must expand through an increase in its employment share $\mu_{o,m}$. 

Depending on the assumptions one is willing to make regarding factor substitution, this model can be taken either as one that makes opposite predictions to the Heckscher-Ohlin model, or one that makes similar predictions to the Heckscher-Ohlin model, but with endogenous labor supply decisions.

In either case, it adds a new margin of analysis: what happens (endogenously) to female-to-male labor supply decisions after foreign demand shocks? In the next section, we test the main prediction of the model using plausibly exogenous variation in Tunisian exports induced by foreign demand shocks and its impact on local labor markets.

\section{Data and Stylized Facts} 

\paragraph{Data} We combine multiple data sources to be able to empirically test the predictions of the theory present in this paper. Labor market indicators come from the Tunisian Government’s National Survey on Population and Employment (ENPE). Among other information, the ENPE reports worker-level employment status by industry (if any), gender, and region for 2006, 2008, 2009, 2010, 2011, 2013, 2015, and 2016\footnote{Specifically, we have data available for eight years in that period but in creating our instrument we have difference and lag the data, reducing the sample.}.  Our strategy is to compare the changes between each pair of years and report the results that emerge from pooling all year-pairs. 

Trade data come from two sources. Merchandise trade data come from UNCOMTRADE while service trade numbers come from the WTO-OECD BaTiS database\footnote{UNCOMTRADE data can be retrieved from \url{http://comtrade.un.org} while WTO-OECD BaTiS can be downloaded from \url{https://www.wto.org/english/res_e/statis_e/trade_datasets_e.htm}.}. We then used concordance tables to map HS and BOPS trade product codes onto ISIC industry codes using concordance tables from WITS. Finally, we mapped ISIC industry codes onto Tunisian Economic Activity Nomenclature (NAT), which is a domestic economic activity classification based on NACE Revision 3. The resulting data have 21 industries and 24 regions.  We were then able to construct a provincial panel for labor market and trade indicators. For each region, we can trace the evolution of employment by gender and sector as well as the district-level exposure to foreign demand shocks, which are shown later in this section.  Additional details on data construction can be found in Appendix \ref{appendix: data}.

As a benchmark, we present in Table \ref{table:summary-stats} the summary statistics of the main variables included in the baseline regressions. Exports and employment are typically growing but on average the change in the female-to-male employment ratio is close to zero. Importantly, the support of the female employment growth distribution is as wide as the male employment growth and their standard deviations are also quite similar \textemdash meaning that the dispersion of changes in employment can be similarly wide across genders. We also present further summary statistics, including the initial and final distributions of employment across regions and industries, in Appendix \ref{appendix: summary}.

\begin{table}[htp]

\resizebox{\textwidth}{!}{

\begin{tabular}{llllll}
\hline
Variable                                          & N   & Mean   & Std. dev. & Min      & Max     \\ \hline
    Change in Exports Exposure   (billion USD)        & 120 & -.052  & .374      & -1.546   & 1.376   \\
    Change in Female-to-Male Employment   Ratio       & 120 & -.004  & .072      & -.386    & .277    \\
    Change in Female Employment                       & 120 & 592.7  & 6152.9    & -20144.3 & 17400.2 \\
    Change in Male Employment                         & 120 & 1592.5 & 5740.2    & -12011.2 & 15507.2 \\
    Change in Education High School or   Higher Share & 120 & -.008  & .014      & -.048    & .023    \\
    Change in Urban Share                             & 120 & -.0002 & .063      & -.248    & .267    \\ \hline
\end{tabular}
}
\caption{\textbf{Summary Statistics}. \footnotesize{Authors' calculations with data from Tunisian Government’s National Survey on Population and Employment (ENPE), UNCOMTRADE, and WTO-OECD BaTiS.} } \label{table:summary-stats}
\end{table}

\paragraph{Stylized facts} Low female labor-force participation rates and high female unemployment rates characterize Tunisian labor markets. While Tunisian female labor force participation (29\%) is higher than other Middle Eastern and Northern Africa (MENA) countries (20\%), it is about half of the rate of OECD countries (65\%) and much lower than other developing regions such as Latin America (55\%) and Sub-Saharan Africa (62\%)\footnote{These statistics are all from the World Bank's World Development Statistics for 2021. \url{https://data.worldbank.org/indicator/SL.TLF.ACTI.FE.ZS?locations=ZQ-ZJ-ZG-OE-TN}.}. The gender gap in unemployment has increased substantially since the 2011 Arab spring. Female unemployment rates have never converged back to pre-Arab spring levels, as seen in Table \ref{table:unemp}. 

\begin{table}[htp]

\resizebox{\textwidth}{!}{
\begin{tabular}{llllllllllll}
\hline
                             & \textbf{2006} & \textbf{2008} & \textbf{2009} & \textbf{2010} & \textbf{2011} & \textbf{2013} & \textbf{2015} & \textbf{2016} \\ \hline
 \textbf{Male unemployment} & 11.5\% & 11.1\% & 11.3\% & 10.9\% & 15.0\% & 13.3\% & 12.4\% & 12.4\% \\ 
 \textbf{Female unemployment} & 15.1\% & 15.9\% & 18.8\% & 18.9\% & 27.4\% & 23.0\% & 22.2\% & 23.5\% \\ \hline
\end{tabular}
}
\caption{\footnotesize{Authors' calculations with data from Tunisian Government’s National Survey on Population and Employment (ENPE).} } \label{table:unemp}
\end{table}

Tunisia is of particular interest because it has a large degree of gender segmentation in labor markets, but female workers are a non-negligible part of labor markets. In some MENA countries, female access to labor markets is very limited, such that there is limited variability in female response to labor market shocks. To test the hypothesis described above, one needs both relevant female participation in labor markets and a high degree of gender segmentation in labor markets.

In a very broad sense, as we alluded to in the previous section, the two largest tradable sectors that are relatively female intensive lost relative importance over the sample period. Apparel (HS61 + HS62) and electrical machinery (HS 85), decreased their shares in total  merchandise exports in 18pp in the period under analysis, as shown in Table \ref{table:apparel}.

\begin{table}[htp]

\resizebox{\textwidth}{!}{
\begin{tabular}{llllllllllll}
\hline
                             & \textbf{2006} & \textbf{2007} & \textbf{2008} & \textbf{2009} & \textbf{2010} & \textbf{2011} & \textbf{2012} & \textbf{2013} & \textbf{2014} & \textbf{2015} & \textbf{2016} \\ \hline
\textbf{Apparel (HS61 + HS62)}        & 26\% & 23\% & 19\% & 21\% & 19\% & 18\% & 16\% & 16\% & 16\% & 15\% & 16\% \\
\textbf{Electrical machinery (HS 85)} & 20\% & 18\% & 14\% & 15\% & 14\% & 13\% & 12\% & 12\% & 12\% & 11\% & 12\% \\ 
\textbf{Other merchandise exports}                & 55\% & 59\% & 66\% & 63\% & 68\% & 68\% & 73\% & 72\% & 72\% & 74\% & 73\% \\ \hline
\end{tabular}
}
\caption{\footnotesize{\textbf{Tunisia: Distribution of Merchandise Exports, by Industry}. Authors' calculations with UNCOMTRADE data. Calculated as exports reported by Tunisia having as destination all countries in the world.}} \label{table:apparel}
\end{table}

\begin{figure}[htp!]
    \centering
    \includegraphics[width=0.8\textwidth]{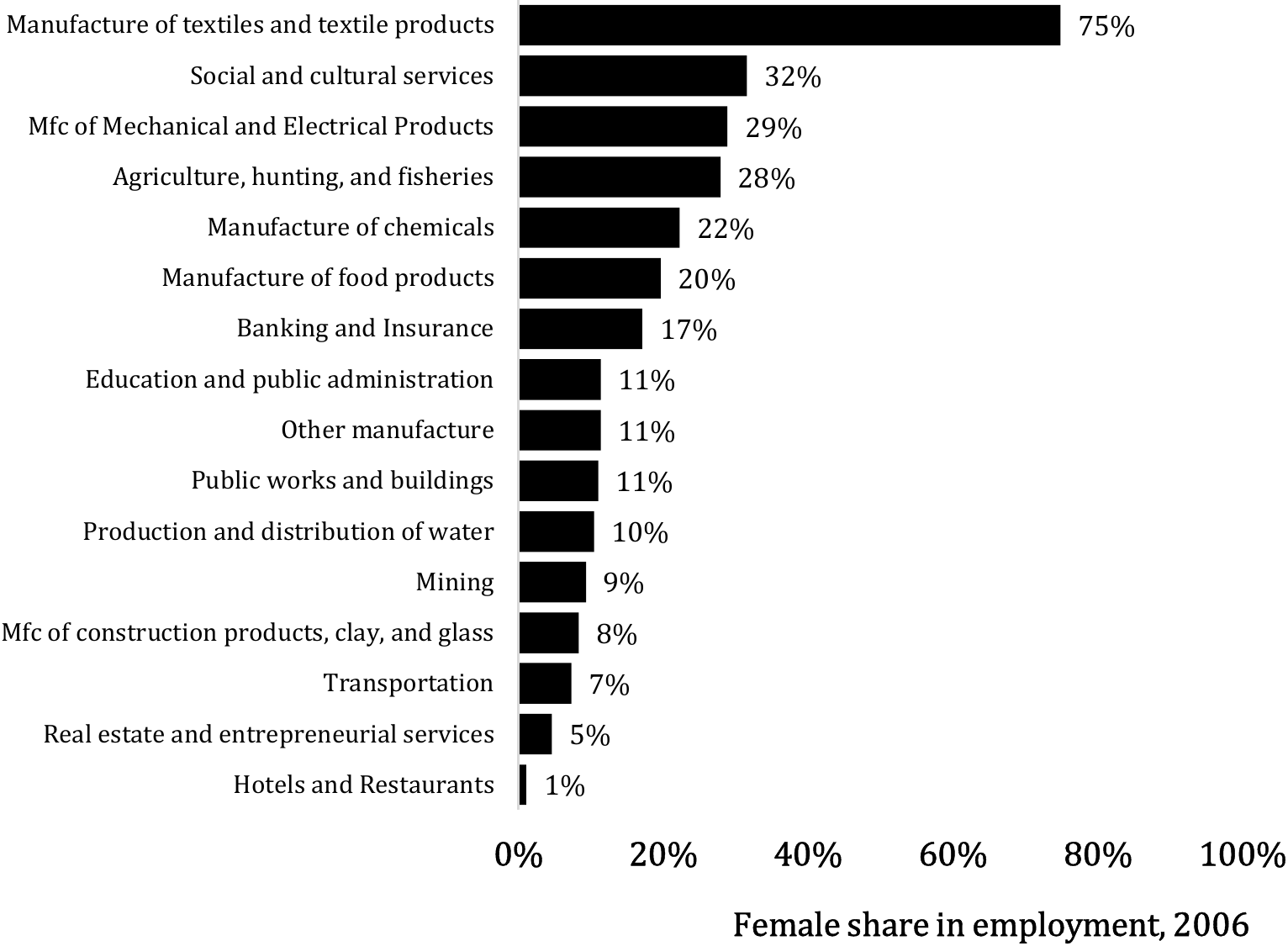}
    \caption{\textbf{Gender composition across industries}. Distribution of female intensity across industries at the beginning of the period (2006). }
    \label{fig:distribution-industries}
\end{figure}

\begin{figure}[htp!]
    \centering
    \includegraphics[width=0.6\textwidth]{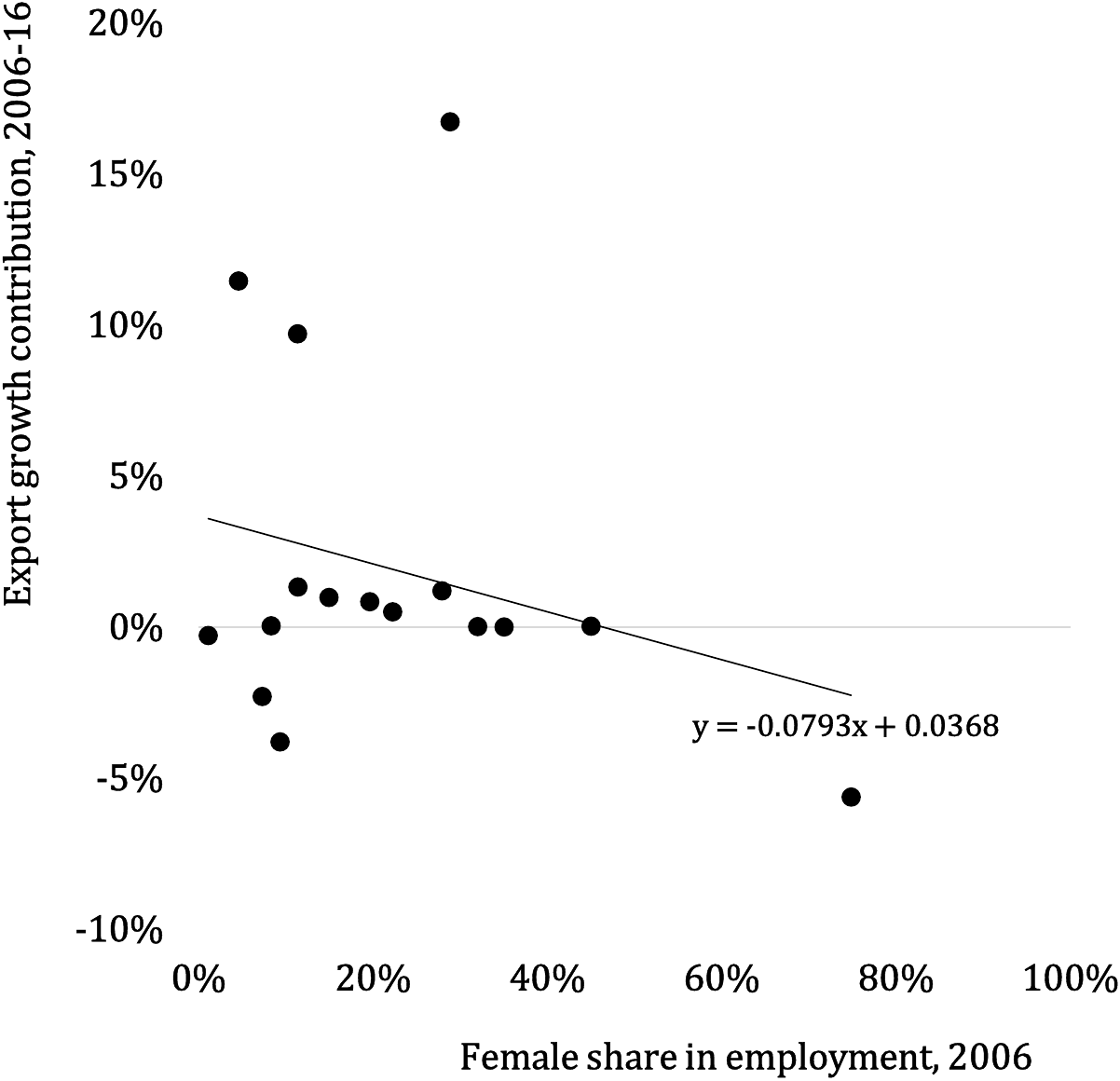}
    \caption{\textbf{Gender composition and contribution to export growth across industries}. Distribution of female intensity across industries at the beginning of the period The Figure shows that industries that contributed the most to aggregate export growth (2006-16) tended to be male-intensive in the base year (2006). }
    \label{fig:growth-industries}
\end{figure}

In order to go deeper into these facts, we separate the industry composition of Tunisia into 15 different industries and plot their initial female-intensity in Figure \ref{fig:distribution-industries}. In 2006, the female share of employment across different industries ranged anywhere from 1\% to 75\%. The most female-intensive industries are garments, social and cultural services, education, and manufacturing of mechanical and electrical products. As as seen on Figure \ref{fig:growth-industries}, industry contribution to export growth in the 2006-16 period is negatively correlated with initial female intensity, meaning that, on average, the most female intensive industries contributed the least to exports growth in Tunisia over the sample period. We also show the underlying data for Figure \ref{fig:growth-industries} on Table \ref{table:growth-contribution} in Appendix \ref{appendix: summary}.

Figure \ref{fig:growth-industries} also implies that, in Tunisia, industries with different female intensities have faced differential growth in exports. As illustrated by the theoretical model, this is a necessary condition to assess this causal mechanism. 

\begin{figure}[htp!]
    \centering
    \includegraphics[scale=0.45]{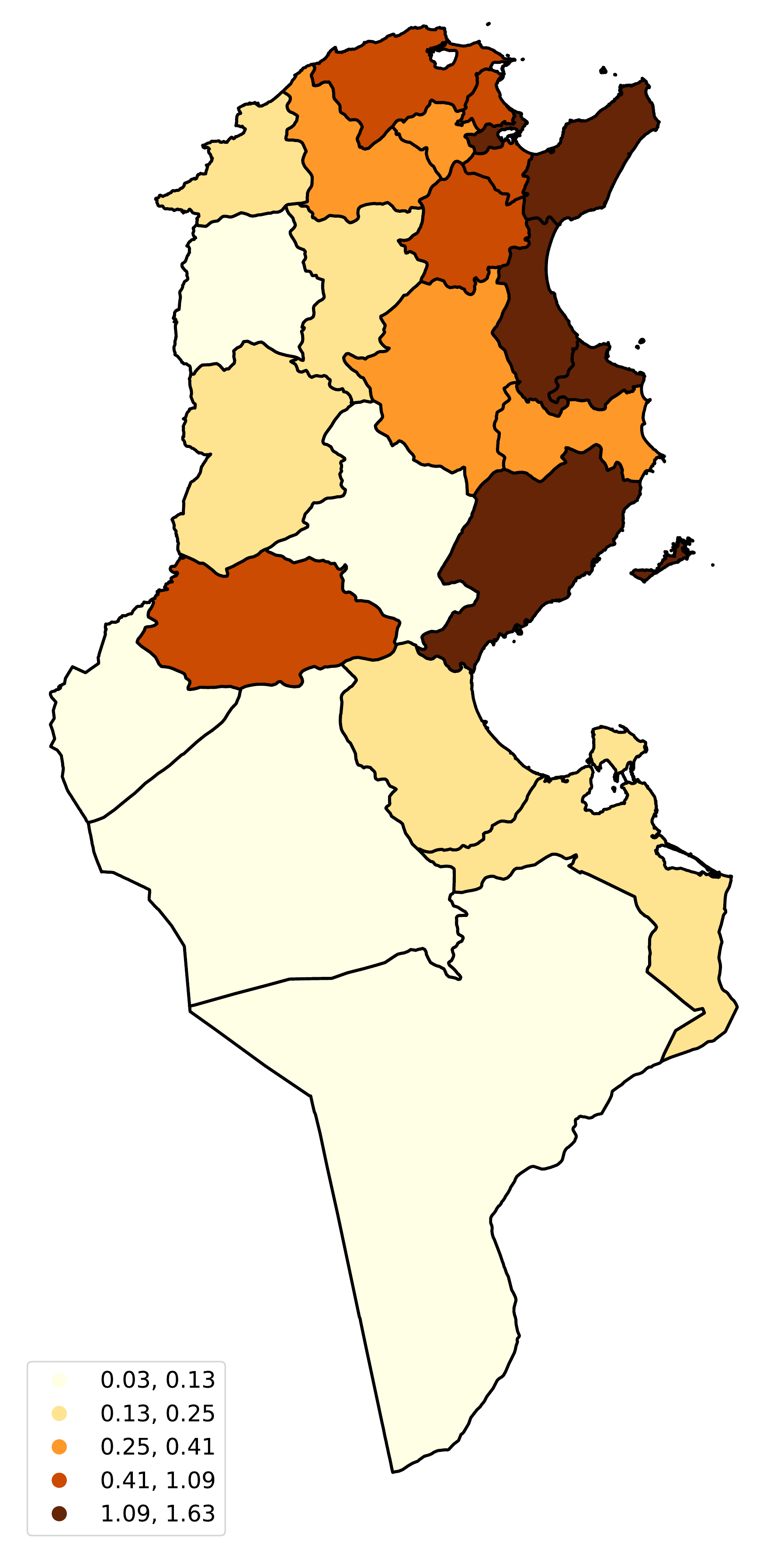}
    \caption{\textbf{District-Level Exposure to Exports, 2016} This Figure denotes a district-specific exposure to exports, defined as $\widetilde{X}_{r,t} \equiv \sum_{i\in\mathcal{I}}\frac{L_{r,i,t-1}}{L_{r,t-1}}\cdot{\widetilde{X}}_{i,t}$,  where ${\widetilde{X}}_{i,t}$ denotes total exports of industry $i$ at period $t$; $L_{r,i,t}$ denotes total employment in region $r$ and industry $i$; and $L_{i,t}\equiv\sum_{r\in\mathcal{R}}\ L_{r,i,t}$ is total aggregate employment in industry $i$. Period $t$ is taken as the base year.}
    \label{fig:map}
\end{figure}

We also leverage variation in exposure to exports and labor force composition across Tunisian regions. The Northeast and Mid-East regions specialize in manufacturing, while the cities in the Grand Tunis region primarily specialize in services. While we do not directly observe exports by district, we leverage the regional variation in labor force across industries and variation in exports across industries to calculate a district-level index of of exposure to exports, as shown in Figure \ref{fig:map}. Details on the methodology follow in the next section.

In terms of destination of exports, European markets dominate most of exports, with France (32\%), Italy (22\% - 17\%), and Germany (8\% - 11\%)  being the largest destination markets during this time period. We report the distribution of merchandise exports over in Table \ref{table:destination-exports} in Appendix \ref{appendix: summary}.

\newpage

\section{Empirical Strategy and Results}

\paragraph{Method} In order to measure the impact of exports over local labor markets, we would ideally observe the share of exports produced in each region $r\in\mathcal{R}$. These data, however, are typically not available for any country. Even for countries that do report export by region, these data typically do not account for production location, but rather the location of the exporting firm – which could be an intermediary. 

In order to circumvent this limitation, we assign each local labor market a share of total exports of a given industry based on that local labor market's share of the given industry's total employment in the period preceding the measured growth in exports. Here we calculate the change between each pair of years, pool these changes, and then report the pooled results. Each change between $t-1$ and $t$ should be interpreted as the next available period in the sample. We formally define local labor market exposure to export growth as:
\begin{equation*}
    \Delta{\widetilde{X}}_{r,t}\equiv\sum_{i\in\mathcal{I}}\frac{L_{r,i,t-1}}{L_{r,t-1}}\cdot\Delta{\widetilde{X}}_{i,t}=\sum_{i\in\mathcal{I}}\frac{L_{r,i,t-1}}{L_{r,t-1}}\cdot\left({\widetilde{X}}_{i,t}-{\widetilde{X}}_{i,t-1}\right)  
\end{equation*}

\noindent in which ${\widetilde{X}}_{i,t}$ denotes total exports of industry $i$ at period $t$; $L_{r,i,t}$ denotes total employment in region $r$ and industry $i$; and $L_{r,t}\equiv\sum_{i\in\mathcal{R}}\ L_{r,i,t}$ is total aggregate employment in region $r$. This kind of shift-share approach mirrors the approach of many papers that study the impact of trade shocks on  local labor markets, including \textcite{autor_china_2013} and \textcite{dix-carneiro_trade_2015}, for imports exposure, and \textcite{robertson_international_2021} for exports exposure. As we note in more detail in the caveats subsection, this approach compares regions that are differentially exposed to exports and can only capture \textit{relative effects} rather than identifying country-wide effects. \textcite{muendler_trade_2017} reviews the literature of the recent methods. 

Table \ref{table:industries} in Appendix \ref{appendix: summary} shows the industries \textit{i} used in the analysis.  Given the shares $\frac{L_{r,i,t}}{L_{r,t}}$, we can potentially estimate the effect of exports over local labor markets by regressing some variable of interest $\Delta O_{r,i,t+1}$, which in our case will be primarily the change in the female-to-male employment ratio, on the shift-share regressand above, provided that the shifters $\Delta{\widetilde{X}}_{i,t}$ are as good as random (for a formal treatment, see \cite{borusyak_quasi-experimental_2020}). If that were the case, we would be able to run the regression:

\begin{equation}
    \Delta O_{r,t+1}=\alpha+\beta\Delta{\widetilde{X}}_{r,t} + {\boldsymbol{Z}'}_{r,t} \boldsymbol{\delta} +\varepsilon_{r,t+1}
\end{equation}

\noindent for which estimation of $\beta$ is consistent if $E\left[\Delta{\widetilde{X}}_{i,t}\cdot\varepsilon_{r,t+1}\middle|{\ \boldsymbol{Z}}_{r,t},\ L_{r,i,t-1}\right]=0$ for every $i$ and $r$ pair. That is, if conditional on controls ${\boldsymbol{Z}}_{r,i,t}$ and on shares, changes in exports are uncorrelated with unobserved local labor markets shocks. 

Since exports depend partially on domestic human capital and technology use, which can be correlated with characteristics of local labor markets, the shifters $\Delta{\widetilde{X}}_{i,t}$ are likely not exogenous. For that reason, we instrument $\Delta{\widetilde{X}}_{r,i,t}$ with increases in foreign demand.  Foreign demand is measured by changes in dollar GDP in foreign destinations, as in our model.

For clarity, we need to introduce some notation. Recall that $\mathcal{S}$ is set of all regions in the world (both foreign countries and Tunisian regions). Denote $r \in \mathcal{T}$ the set of regions in Tunisia. Tunisia exports to countries other than itself \textemdash or to destinations $d\in\mathcal{S}\setminus{\mathcal{T}}$. We denote the exports of each industry $i$ as the sum of its sales to every foreign destination: ${\widetilde{X}}_{i,t}=\sum_{d\in\mathcal{S}\setminus{\mathcal{T}}}\widetilde{X}_{d,i,t}$.   Our foreign-demand instrument leverages the correlation between changes in exports to destination $d$ and changes in dollar GDP, which is expected given the gravity structure typical of international trade. It is:

\begin{equation}\label{eq: instrument}
    \Delta{\bar{X}}_{r,t}\equiv\sum_{i\in\mathcal{I}}\frac{L_{r,i,t-1}}{L_{r,t-1}}\cdot\sum_{d\in\mathcal{S}\setminus{\mathcal{T}}}\frac{\widetilde{X}_{d,i,t-1}}{{\widetilde{X}}_{i,t-1}}\cdot\Delta Y_{d,t}    
\end{equation}

\noindent where $\frac{\widetilde{X}_{d,i,t}}{{\widetilde{X}}_{i,t}}$ denotes country $d$’s share of industry $i$’s exports; and $\Delta Y_{d,t}$ is the change in U.S. dollar GDP in country $d$. In other words, there is some common vector of foreign demand shocks (changes in foreign GDP) but Tunisian industries are, in each period, differentially exposed to foreign markets proportionately to their export shares to foreign markets in the previous period. In turn, each region in Tunisia is differentially exposed to each industry through its labor markets and hence inherit some differential exposure to foreign demand shocks.

Note that this takes into account \textbf{every country that Tunisia exports to} in \textbf{every industry}. Therefore, this instrumentation strategy is similar to those such as in \textcite{aghion_heterogeneous_2022}, in which one takes demand in all foreign destinations as a proxy for exogenous variation in demand in a given industry.  The variation in foreign GDP is assumed to be driven by factors outside of Tunisia and therefore isolate the exogenous changes in export demand.

Estimation now takes the form of two-stage least squares.  In the first stage, we estimate the change in Tunisian exports that are explained by changes in the GDP in the importing country.  Specifically, the first stage is:

\begin{equation}\label{eq: reduced-form}
    \Delta{\widetilde{X}}_{r,t+1}=\omega+\gamma\Delta{\bar{X}}_{r,t}\ +\ {\boldsymbol{Z}'}_{r,t} \boldsymbol{\phi} +{\bar{\varepsilon}}_{r,t+1}
\end{equation}

\noindent and the second stage is:

\begin{equation}
    \Delta O_{r,t+1}=\alpha+\beta\Delta{\hat{X}}_{r,t}\ +\ {\boldsymbol{Z}'}_{r,t} \boldsymbol{\delta} +\varepsilon_{r,t+1}
\end{equation}

\noindent where $\Delta{\hat{X}}_{r,t}$ are the predicted values of the first stage regression. Now estimation of $\beta$ is consistent if $E\left[\Delta Y_{d,t}\cdot\varepsilon_{r,t}\middle|{\ \boldsymbol{Z}}_{r,t}, \widetilde{X}_{d,i,t}, L_{r,i,t-1}\right]=0$ for every $d$ and $r$ pair -- i.e. if changes in foreign demand are uncorrelated with unobserved factors that drive changes local labor markets in Tunisia. 

Tunisia is a small open economy. Therefore, it is unlikely that changes in foreign demand are correlated with unobserved factors that differentially drive changes local labor markets – i.e., this instrument is likely valid. Even though the variables are already first-differenced, to control for the possibility that there could be simultaneous technological shocks that are correlated with foreign GDP shocks, we also include time dummies that absorb common time-specific shocks. We also include district-fixed effects to control for time-invariant unobserved confounders at a growth rate level. Additional control variables include changes in the share of high school population and the changes in the share of urban population.

\begin{figure}[htp!]
    \centering
    \includegraphics[width=0.65\textwidth]{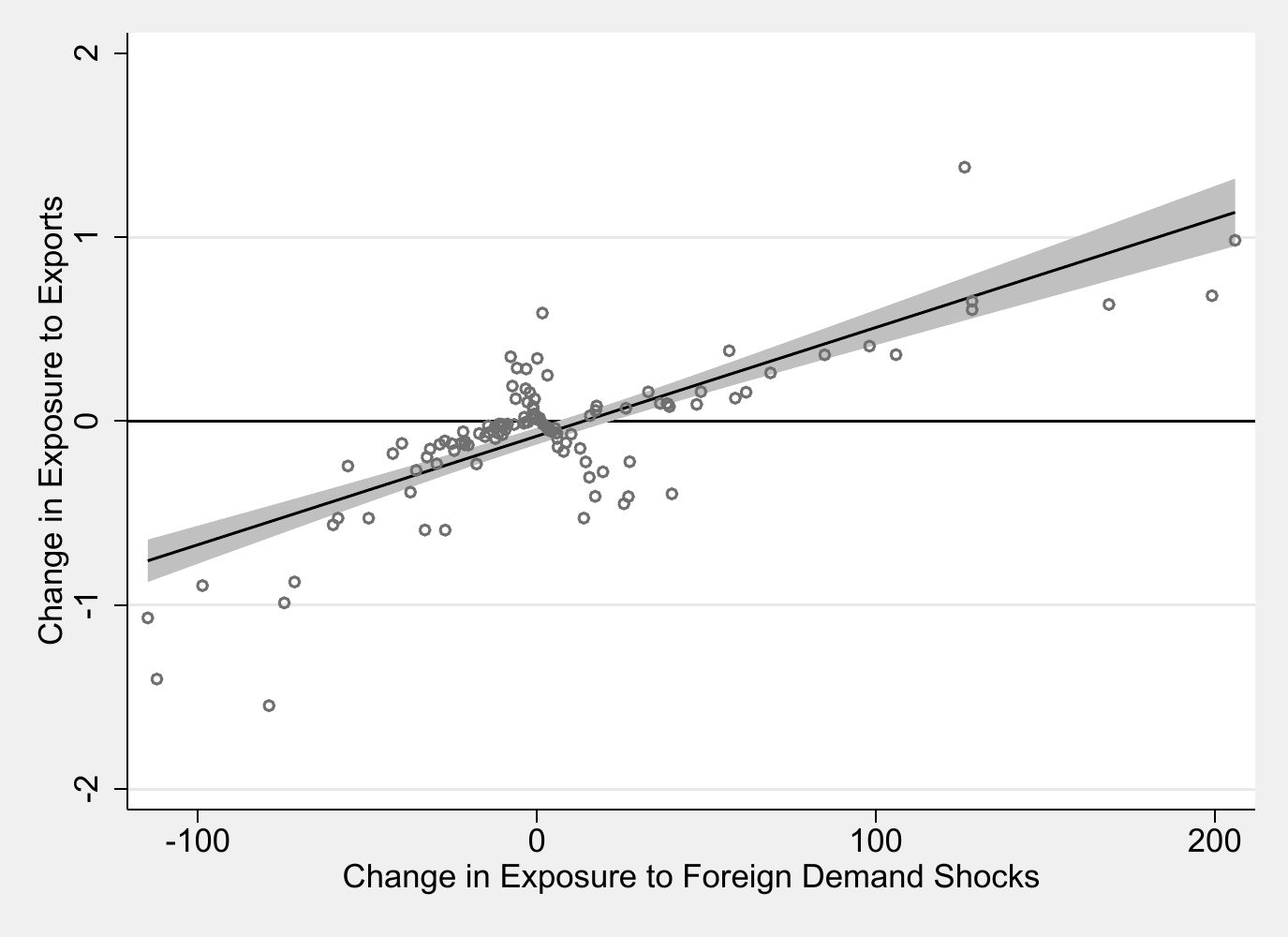}
    \caption{\textbf{First-Stage Regression}. This Figure shows the correlation between the instrument defined in equation \eqref{eq: instrument} and the left-hand side of equation \eqref{eq: reduced-form}.}
    \label{fig:instrument}
\end{figure}

Figure \ref{fig:instrument} shows the correlation between the instrument defined in equation \eqref{eq: instrument} and the left-hand side of equation \eqref{eq: reduced-form}. Furthermore, as changes in exposure to exports are strongly correlated with changes in exposure to foreign demand shocks ($f-stat > 130$), the instrument is relevant. Satisfying the exclusion restriction and instrument relevance, we can appropriately interpret the results in this section as the causal effect of exports on local labor markets in Tunisia during this time period.

\paragraph{Results}  Since the change in exports was higher in male-intensive industries, the theoretical mechanism described above predicts that the female-to-male employment ratio should decline. The empirical estimates confirm the theoretical prediction. Panel A of Table \ref{table:reg1} shows the results of the two-stage least squared regression of the change in female-to-male employment ratio on the change in exposure to exports instrumented by the change in exposure to foreign demand. Our preferred specification, which includes time- and region-fixed-effects as well socio-demographic controls (lagged change urban share and change high school share of the region), is column (3). 

All of our estimates are on differenced variables without log transformations. As the female-to-male employment ratio is a ratio, we can interpret the coefficient in terms of percentage points. Conversely, the interpretation on the coefficient on male and female employment is in terms on the number of jobs. To better understand the economic magnitude of these results, we also provide the normalized coefficients by using the standard deviations of both variables.

A 1 billion USD increase in exposure to exports led to an average decrease of 6.8pp in the female-to-male employment ratio, which is statistically significant at the 5\% confidence level and in the same direction as predicted by the theoretical mechanism.  An increase in exposure to exports of 1 standard deviation decreases the female-to-male employment ratio by 0.137 standard deviations.

We can decompose this result by running separate regressions of female and male employment, respectively, on exposure to exports. We present these results on Panels B and C of Table \ref{table:reg1}. This decomposition shows that, in fact, the effects of increased exports induced by foreign demand shocks have opposite signs on female and male employment.

While the point estimate for female employment is of 7,903 \textit{fewer} female jobs in response to an increase of export exposure of 1 billion USD, the point estimate for male employment is of 2,418 \textit{additional} jobs. Only the coefficient for females is statistically significant. These results are in line with the findings of \textcite{kongar_importing_2006}, whereby trade integration reduces the gender wage gap through a reduced demand for female labor.

Normalizing in terms of standard deviations highlights that female employment is more responsive to trade at the extensive margin. An increase in exposure to exports of 1 standard deviation decreases the female employment ratio by 0.062 standard deviations while it increases male employment by 0.009 standard deviations.

No increase in male employment in response to the foreign demand shock is also consistent with the theoretical model. The model only makes a prediction regarding the female-to-male employment ratio, not the independent move about either of those equilibrium objects. In the model, the household is optimizing the opportunity cost of female labor supply relative to male labor supply, given their marginal benefit (wages). 

One final point that should be stressed is that these results compare regions more exposed to foreign demand shocks to those relatively less exposed to shocks. Therefore, without additional assumptions, results should be interpreted as relative effects with distributional implications across regions of Tunisia rather than aggregate results. We have a richer discussion about aggregation on the caveats subsection.

\begin{table}[htp!]\label{table:reg1}
\centering
\begin{tabular}{lccc}
\hline
\multicolumn{4}{l}{Panel   A. Response Variable: Change in Female-to-Male Employment Ratio}             \\ \hline
                                           & (1)               & (2)               & (3)                \\
Change in Exports Exposure (in billion   USD) &
  \begin{tabular}[c]{@{}l@{}}0.002\\    \footnotesize{(0.018)}\end{tabular} &
  \begin{tabular}[c]{@{}l@{}}-0.068*\\  \footnotesize{(0.035)}\end{tabular} &
  \begin{tabular}[c]{@{}l@{}}-0.068**\\    \footnotesize{(0.031)}\end{tabular} \\ \hline
\multicolumn{4}{l}{Panel   B. Response Variable: Change in Female Employment}                           \\ \hline
                                           & (1)               & (2)               & (3)                \\
Change in Exports Exposure   (billion USD) &
  \begin{tabular}[c]{@{}l@{}}-1,929.460\\    \footnotesize{(2,257.502)}\end{tabular} &
  \begin{tabular}[c]{@{}l@{}}-8,173.887**\\    \footnotesize{(3,455.325)}\end{tabular} &
  \begin{tabular}[c]{@{}l@{}}-7,903.951***\\    \footnotesize{(3,048.396)}\end{tabular} \\ \hline
\multicolumn{4}{l}{Panel   C. Response Variable: Change in Male Employment}                             \\ \hline
                                           & (1)               & (2)               & (3)                \\
Change in Exports Exposure   (billion USD) &
  \begin{tabular}[c]{@{}l@{}}1,756.212\\    \footnotesize{(1,842.006)}\end{tabular} &
  \begin{tabular}[c]{@{}l@{}}1,140.393\\    \footnotesize{(3,026.194)}\end{tabular} &
  \begin{tabular}[c]{@{}l@{}}2,418.129\\    \footnotesize{(3,133.433)}\end{tabular} \\ \hline
\multicolumn{4}{l}{Panel   D. First-Stage. Response Variable: Change in Exports Exposure (billion USD)} \\ \hline
                                           & (1)               & (2)               & (3)                \\
Change in Foreign Demand Exposure   (billion USD) &
  \begin{tabular}[c]{@{}l@{}}.0058***\\    \footnotesize{(.0006)}\end{tabular} &
  \begin{tabular}[c]{@{}l@{}}.0046***\\    \footnotesize{(.0008)}\end{tabular} &
  \begin{tabular}[c]{@{}l@{}}.0045***\\    \footnotesize{(.0004)}\end{tabular} \\
F-statistic                                & 77.82             & 30.25             & 139.95             \\ \hline
Time   Fixed-Effects                       &                   & \checkmark        & \checkmark                  \\
Region Fixed-Effects                     &                   & \checkmark        & \checkmark                  \\
Socio-Demographic   Controls               &                   &                   & \checkmark                  \\
N                                          & 120               & 120               & 120                \\ \hline

\multicolumn{4}{c}{ \parbox{\columnwidth}{ \footnotesize{ Notes: Region cluster robust standard errors in parenthesis. * $p<0.1$; ** $p<0.05$; *** $p<0.01$. Demographic controls include the change in the share of high-school population and the change in the share of urban population. Years available in the survey are: 2006, 2008, 2009, 2010, 2011, 2013, 2015, and 2016. In creating our instrument we had to difference and lag the data, reducing the sample. } } } \\

\end{tabular}

\caption{Baseline Employment and Employment Ratio Regressions}

\end{table}

\newpage

\paragraph{Mechanism} 
The relative contraction of the female-intensive industry frees up women and men.  Our estimates suggest that men shift from the contracting industry to the expanding industry, leading to a slightly positive, but statistically insignificant, increase in employment for men.  Women, however, cannot shift easily into the male-intensive industry, so they simply drop out of the labor force resulting in a statistically signficant drop in female employment.   Interpreted through the lens of the theoretical model, this suggests that households may be substituting male for female labor supply. ENPE has no information on wages, so one cannot confirm the substitution of lower wage female jobs for higher wage male jobs directly. Nonetheless, estimating the effect over some other margins can shed some light into the issue.

For instance, we first estimate the effect of the change in exposure to exports on the change of female and male unemployment, respectively. Neither estimate is statistically significant, but the results indicate that, on average, a 1 billion USD increase in export exposure displaced 272 women into unemployment and 1280 males out of unemployment. These results in Panels A and B of Table \ref{table:reg2}.

Therefore, the induced change on female unemployment is either zero, negative or, if positive, more than one order of magnitude smaller than changes in employment. Combined with the results from the regressions on employment, this shows that the observed foreign demand shocks, known to be concentrated on male-intensive industries, primarily induced females to move out of the labor force rather than into unemployment.

This result is consistent with households optimizing quantities of female and male labor supply in a gender-segmented labor market. If there were large increases in the unemployment margin induced by foreign-demand shocks, then any analytical exercise could not abstract away from modeling involuntary unemployment. 

More evidence comes from comparing the response of married versus single female workers. If households are indeed substituting male for female labor supply, then one would expect the effect to be stronger among married women. This is indeed the case. As shown in Panels C and D of Table \ref{table:reg2}, most of the variation in female employment following a foreign demand shock comes from married female workers. 

On average, a 1 billion USD increase in export exposure led to 4,605 fewer jobs among married female workers and 2,501 fewer jobs among single female workers. Furthermore, the results are statistically significant at the 5\% confidence level for married female workers and statistically insignificant for female workers, indicating that the relationship is much tighter for married women. This result is also suggestive evidence that the empirical response is consistent with the theoretical mechanism discussed before. 

Importantly, we see no similar differences when comparing the effect across single and married male workers. The effect is close to zero and statistically insignificant for both groups. We take this as evidence that these shocks not only differentially impact both groups, but differentially impact married and single workers \textemdash with an intuition similar to a ``triple differences'' design to policy changes: not only trade impacts man and women differentially, it impacts them differentially through the channel of marriage.

Both the primary result on the female-to-male employment ratio and the secondary results presented above are in line with the predictions of the theoretical model. The model thus is consistent with the empirical evidence in Tunisia -- and its predictions can be tested for other countries.

\begin{table}[htp]\label{table:reg2}
\begin{tabular}{lccc}
\hline
\multicolumn{4}{l}{Panel   A. Response Variable: Change in Unemployed Females}                          \\ \hline
                                           & (1)                & (2)               & (3)               \\
Change in Exports Exposure   (billion USD) &
  \begin{tabular}[c]{@{}l@{}}-1,387.8\\    \footnotesize{(1,617.9)}\end{tabular} &
  \begin{tabular}[c]{@{}l@{}}147.3\\    \footnotesize{(2,611.8)}\end{tabular} &
  \begin{tabular}[c]{@{}l@{}}272.0\\    \footnotesize{(2,749.1)}\end{tabular} \\ \hline
\multicolumn{4}{l}{Panel   B. Response Variable: Change in Unemployed Males}                            \\ \hline
                                           & (1)                & (2)               & (3)               \\
Change in Exports Exposure   (billion USD) &
  \begin{tabular}[c]{@{}l@{}}-1,750.4\\    \footnotesize{(1,680.8)}\end{tabular} &
  \begin{tabular}[c]{@{}l@{}}-602.6\\    \footnotesize{(2,243.9)}\end{tabular} &
  \begin{tabular}[c]{@{}l@{}}-1,279.6\\    \footnotesize{(2,156.9)}\end{tabular} \\ \hline
\multicolumn{4}{l}{Panel   C. Response Variable: Change in Female Employment, Married Women}            \\ \hline
                                           & (1)                & (2)               & (3)               \\
Change in Exports Exposure   (billion USD) &
  \begin{tabular}[c]{@{}l@{}}-343.177\\    \footnotesize{(1,348.004)}\end{tabular} &
  \begin{tabular}[c]{@{}l@{}}-4,540.806**\\    \footnotesize{(1,993.872)}\end{tabular} &
  \begin{tabular}[c]{@{}l@{}}-4,605.166**\\    \footnotesize{(1,897.571)}\end{tabular} \\ \hline
\multicolumn{4}{l}{Panel   D. Response Variable: Change in Female Employment, Single Women}             \\ \hline
                                           & (1)                & (2)               & (3)               \\
Change in Exports Exposure   (billion USD) &
  \begin{tabular}[c]{@{}l@{}}-1,446.643\\    \footnotesize{(1,127.362)}\end{tabular} &
  \begin{tabular}[c]{@{}l@{}}-2,750.395*\\    \footnotesize{(1,549.210)}\end{tabular} &
  \begin{tabular}[c]{@{}l@{}}-2,501.416\\    \footnotesize{(1,429.758)}\end{tabular} \\ \hline
\multicolumn{4}{l}{Panel   E. Response Variable: Change in Male Employment, Married Men}            \\ \hline
                                           & (1)                & (2)               & (3)               \\
Change in Exports Exposure   (billion USD) &
  \begin{tabular}[c]{@{}l@{}}1,605.056\\    \footnotesize{(1,050.573)}\end{tabular} &
  \begin{tabular}[c]{@{}l@{}}650.071\\    \footnotesize{(1,482.831)}\end{tabular} &
  \begin{tabular}[c]{@{}l@{}}301.866\\    \footnotesize{(1,454.947)}\end{tabular} \\ \hline
\multicolumn{4}{l}{Panel   F. Response Variable: Change in Male Employment, Single Men}             \\ \hline
                                           & (1)                & (2)               & (3)               \\
Change in Exports Exposure   (billion USD) &
  \begin{tabular}[c]{@{}l@{}}-162.849\\    \footnotesize{(1,222.424)}\end{tabular} &
  \begin{tabular}[c]{@{}l@{}}164.996\\    \footnotesize{(1,464.048)}\end{tabular} &
  \begin{tabular}[c]{@{}l@{}}1,756.482\\    \footnotesize{(1,421.429)}\end{tabular} \\ \hline
\multicolumn{4}{l}{Panel   G. First-Stage. Response Variable: Change in Exports Exposure (billion USD)} \\ \hline
                                           & (1)               & (2)               & (3)                \\
Change in Foreign Demand Exposure   (billion USD) &
  \begin{tabular}[c]{@{}l@{}}.0058***\\    \footnotesize{(.0006)}\end{tabular} &
  \begin{tabular}[c]{@{}l@{}}.0046***\\    \footnotesize{(.0008)}\end{tabular} &
  \begin{tabular}[c]{@{}l@{}}.0045***\\    \footnotesize{(.0004)}\end{tabular} \\
F-statistic                                & 77.82             & 30.25             & 139.95             \\ \hline
Time   Fixed-Effects                       &                   & \checkmark        & \checkmark                  \\
Region Fixed-Effects                     &                   & \checkmark        & \checkmark                  \\
Socio-Demographic   Controls               &                   &                   & \checkmark                  \\
N                                          & 120               & 120               & 120                \\ \hline

\multicolumn{4}{c}{ \parbox{\columnwidth}{ \footnotesize{ Notes: Region cluster robust standard errors in parenthesis. * $p<0.1$; ** $p<0.05$; *** $p<0.01$. Demographic controls include the change in the share of high-school population and the change in the share of urban population. Years available in the survey are: 2006, 2008, 2009, 2010, 2011, 2013, 2015, and 2016. In creating our instrument we had to difference and lag the data, reducing the sample. } } } \\
\end{tabular}

\caption{Unemployment Regressions and Married vs Single Regressions}
\end{table}

\newpage

\paragraph{Caveats and Robustness} The main limitation of this study is the lack of data on wages in the Tunisian labor force survey (ENPE). The joint prediction of the theoretical model described in this paper is that foreign demand shocks that concentrate on male-intensive industries should decrease both the female-to-male employment ratio and the female relative wage. Most, if not all, theoretic and empirical studies support this result. \textcite{dhyne_foreign_2022} shows that in a large class of models there is a positive pass-through of foreign demand shocks to labor (except when labor is fixed by assumption). \textcite{aguayo-tellez_did_2010} shows that access to liberalized markets through NAFTA in Mexico led to an increase in relative female wages due to positive shocks to the clothing sector. Similarly, \textcite{de_hoyos_exports_2012} shows that the expansion of the \textit{maquilas} in Honduras led to an increase of female relative earnings. \textcite{kis-katos_globalization_2018} document an expansion of the female-intensive sector after input-tariff liberalization in Indonesia, a fact that is consistent with the mechanism presented in the model introduced in this paper.

Furthermore, the fact that we were able to identify that, in Tunisia, changes in the female-to-male employment ratio work primarily through changes in the behavior of married women serves as an additional check on the theoretical mechanism since it suggests that the mechanism operates through joint household decisions.

Another caveat relates to aggregation. The empirical strategy developed here relies on exploiting differential regional variation in exposure to exports. As explained by \textcite{chodorow-reich_regional_2020}, however, strategies that rely on this kind of cross-sectional variation only capture the effect of additional exposure to exports relative to other regions\footnote{In fact, \textcite{chodorow-reich_regional_2020} shows that this estimator is analogous to a differences-in-differences estimator with an additional term that captures spill-overs across regions (SUTVA violations). For instance, \textcite{caliendo_trade_2019} show that the negative relative effect captured from the China Trade Shock estimated by \textcite{autor_china_2013} may be positive in general equilibrium. \textcite{adao_general_2019} develop a similar argument and show that spill-overs can have a non-negligible effect of attenuation or amplification of the relative effect.}.  In short, since shift-share regressions can be interpreted as differences-in-differences, they suffer from a ``missing intercept problem'' \parencite{wolf_missing_2021} with respect to general equilibrium effects and can only capture \textit{relative effects}. For that reason, one cannot extrapolate from the cross-sectional to the aggregate effect without imposing additional assumptions.

In this case, one variable that could make aggregation particularly difficult is migration. If households move across regions in response to negative (positive) shocks, they could make relative estimates larger (smaller) than aggregate estimates. In econometric terms, if domestic migration flows respond positively or negatively to these shocks, then, clearly there are general equilibrium responses and the Stable Unit Treatment Value Assumption (SUTVA) does not hold. A key concern, then, for the consistency of our estimator is that migration flows are largely uncorrelated with our foreign demand shocks. We show in Table \ref{table:reg4} that migration flows are indeed largely uncorrelated with our instrument.

The response is only significant in the specification without controls and would net negative migration in response to a positive demand shock, which is economically insensible. In the specifications (2) and (3), which add fixed-effects and demographic controls, there is no response of changes in district population (which we interpret as net migration) after foreign demand shocks.

\begin{table}[htp]
\begin{tabular}{lccc}
\hline
\multicolumn{4}{l}{Panel   A. Response Variable: Change in District Population}                          \\ \hline
                                           & (1)                & (2)               & (3)               \\
Change in Exports Exposure   (billion USD) &
  \begin{tabular}[c]{@{}l@{}}-2,942.413**\\    \footnotesize{(1,360.602)}\end{tabular} &
  \begin{tabular}[c]{@{}l@{}}-2,801.017\\    \footnotesize{(1,963.837)}\end{tabular} &
  \begin{tabular}[c]{@{}l@{}}-1,737.083\\    \footnotesize{(2,230.651)}\end{tabular} \\ \hline
\multicolumn{4}{l}{Panel   B. First-Stage. Response Variable: Change in Exports Exposure (billion USD)} \\ \hline
                                           & (1)               & (2)               & (3)                \\
Change in Foreign Demand Exposure   (billion USD) &
  \begin{tabular}[c]{@{}l@{}}.0058***\\    \footnotesize{(.0006)}\end{tabular} &
  \begin{tabular}[c]{@{}l@{}}.0046***\\    \footnotesize{(.0008)}\end{tabular} &
  \begin{tabular}[c]{@{}l@{}}.0045***\\    \footnotesize{(.0004)}\end{tabular} \\
F-statistic                                & 77.82             & 30.25             & 139.95             \\ \hline
Time   Fixed-Effects                       &                   & \checkmark        & \checkmark                  \\
Region Fixed-Effects                     &                   & \checkmark        & \checkmark                  \\
Socio-Demographic   Controls               &                   &                   & \checkmark                  \\
N                                          & 120               & 120               & 120                \\ \hline

\multicolumn{4}{c}{ \parbox{\columnwidth}{ \footnotesize{ Notes: Region cluster robust standard errors in parenthesis. * $p<0.1$; ** $p<0.05$; *** $p<0.01$. Demographic controls include the change in the share of high-school population and the change in the share of urban population. Years available in the survey are: 2006, 2008, 2009, 2010, 2011, 2013, 2015, and 2016. In creating our instrument we had to difference and lag the data, reducing the sample. } } } \\
\end{tabular}

\caption{Migration Regressions}
\label{table:reg4}

\end{table}

As another robustness exercise, \textcite{adao_general_2019} (henceforth AAE) argue that the shift-share regression designs tend to underestimate standard errors because regions with similar industry shares may have correlated shocks (residuals) that are independent of geographic location.  To address this, they propose a shift-share estimation strategy that adjusts the standard errors for the possible correlation across regions.  We reestimate our baseline results with the AAE standard errors\footnote{Since we only apply and do not modify their approach, we refer the reader to \textcite{adao_general_2019} for additional details.}.  The results are shown in Table \ref{table:reg-robustness}.  The significance of our results do not change much with the AAE standard errors, with our results standing both qualitatively and quantitatively.

\begin{table}[htp!]
\centering
\begin{tabular}{lccc}
\hline
\multicolumn{4}{l}{Panel   A. Response Variable: Change in Female-to-Male Employment Ratio}             \\ \hline
                                           & (1)               & (2)               & (3)                \\
Change in Exports Exposure (in billion   USD) &
  \begin{tabular}[c]{@{}l@{}}0.001\\    \footnotesize{(0.018)}\end{tabular} &
  \begin{tabular}[c]{@{}l@{}}-0.065**\\  \footnotesize{(0.033)}\end{tabular} &
  \begin{tabular}[c]{@{}l@{}}-0.065**\\    \footnotesize{(0.029)}\end{tabular} \\ \hline
\multicolumn{4}{l}{Panel   B. Response Variable: Change in Female Employment}                           \\ \hline
                                           & (1)               & (2)               & (3)                \\
Change in Exports Exposure   (billion USD) &
  \begin{tabular}[c]{@{}l@{}}-1,880.714\\    \footnotesize{(1,693.848)}\end{tabular} &
  \begin{tabular}[c]{@{}l@{}}-7,752.010***\\    \footnotesize{(2,642.674)}\end{tabular} &
  \begin{tabular}[c]{@{}l@{}}-7,406.008***\\    \footnotesize{(2,624.837)}\end{tabular} \\ \hline
\multicolumn{4}{l}{Panel   C. Response Variable: Change in Male Employment}                             \\ \hline
                                           & (1)               & (2)               & (3)                \\
Change in Exports Exposure   (billion USD) &
  \begin{tabular}[c]{@{}l@{}}1,751.023\\    \footnotesize{(1,894.655)}\end{tabular} &
  \begin{tabular}[c]{@{}l@{}}1,280.684\\    \footnotesize{(2,501.516)}\end{tabular} &
  \begin{tabular}[c]{@{}l@{}}2,517.903\\    \footnotesize{(2,535.838)}\end{tabular} \\ \hline
Time   Fixed-Effects                       &                   & \checkmark        & \checkmark                  \\
Region Fixed-Effects                     &                   & \checkmark        & \checkmark                  \\
Socio-Demographic   Controls               &                   &                   & \checkmark                  \\
N                                          & 120               & 120               & 120                \\ \hline

\multicolumn{4}{c}{ \parbox{\columnwidth}{ \footnotesize{ Notes: Region cluster robust standard errors calculated with the \textcite{adao_general_2019} correction in parenthesis. Standard err * $p<0.1$; ** $p<0.05$; *** $p<0.01$. Demographic controls include the change in the share of high-school population and the change in the share of urban population. Years available in the survey are: 2006, 2008, 2009, 2010, 2011, 2013, 2015, and 2016. In creating our instrument we had to difference and lag the data, reducing the sample. } } } \\

\end{tabular}

\caption{Robustness Standard Error Estimates for Baseline Employment and Employment Ratio Regressions}
\label{table:reg-robustness}
\end{table}

Finally, we re-estimate our baseline regressions controlling for the initial share of manufacturing sector in each region. One potential concern here would be that the variability across employment areas would be exclusively attributed to disparities in local specialization within the manufacturing sector, rather than variations in the overall share of the manufacturing sector. The results are qualitatively the same and quantitatively stronger, meaning that the drop in the female-to-male employment ratio is of a larger magnitude the effect is statistically significant at a stricter confidence level. The results and its decomposition are in Table \ref{table:reg-manufacturing-shares}.

\begin{table}[htp!]
\centering
\begin{tabular}{lccc}
\hline
\multicolumn{4}{l}{Panel   A. Response Variable: Change in Female-to-Male Employment Ratio}             \\ \hline
                                           & (1)               & (2)                        \\
Change in Exports Exposure (in billion   USD)  &
  \begin{tabular}[c]{@{}l@{}}-0.068**\\    \footnotesize{(0.031)}\end{tabular} &
  \begin{tabular}[c]{@{}l@{}}-0.078***\\  \footnotesize{(0.030)}\end{tabular} \\ \hline
\multicolumn{4}{l}{Panel   B. Response Variable: Change in Female Employment}                           \\ \hline
                                           & (1)               & (2)                        \\
Change in Exports Exposure   (billion USD)  &
  \begin{tabular}[c]{@{}l@{}}-7,903.951***\\    \footnotesize{(3,048.396)}\end{tabular} &
  \begin{tabular}[c]{@{}l@{}}-9,118.678***\\    \footnotesize{(2,955.888)}\end{tabular}  \\ \hline
\multicolumn{4}{l}{Panel   C. Response Variable: Change in Male Employment}                             \\ \hline
                                           & (1)               & (2)                        \\
Change in Exports Exposure   (billion USD) &
  \begin{tabular}[c]{@{}l@{}}2,418.129\\    \footnotesize{(3,133.433)}\end{tabular} &
  \begin{tabular}[c]{@{}l@{}}1,885.249\\    \footnotesize{(1,938.007)}\end{tabular}  \\ \hline
\multicolumn{4}{l}{Panel   D. First-Stage. Response Variable: Change in Exports Exposure (billion USD)} \\ \hline
                                           & (1)               & (2)                       \\
Change in Foreign Demand Exposure   (billion USD) &
  \begin{tabular}[c]{@{}l@{}}.0045***\\    \footnotesize{(.0004)}\end{tabular} &
  \begin{tabular}[c]{@{}l@{}}.0047***\\    \footnotesize{(.0004)}\end{tabular} \\
F-statistic                               & 139.95             & 108.57             \\ \hline
Time   Fixed-Effects                                          & \checkmark        & \checkmark                  \\
Region Fixed-Effects                                        & \checkmark        & \checkmark                  \\
Socio-Demographic   Controls                                  &   \checkmark                & \checkmark                  \\
Lagged Share of Manufacturing Employment                                  &                   & \checkmark                  \\
N                                                         & 120               & 120                \\ \hline

\multicolumn{4}{c}{ \parbox{\columnwidth}{ \footnotesize{ Notes: Region cluster robust standard errors in parenthesis. * $p<0.1$; ** $p<0.05$; *** $p<0.01$. Demographic controls include the change in the share of high-school population and the change in the share of urban population. Years available in the survey are: 2006, 2008, 2009, 2010, 2011, 2013, 2015, and 2016. In creating our instrument we had to difference and lag the data, reducing the sample. } } } \\

\end{tabular}

\caption{Robustness Baseline Employment and Employment Ratio Regressions Controlling for Lagged Manufacturing Shares}
\label{table:reg-manufacturing-shares}
\end{table}

\section{Conclusion}

This paper focuses on how gender segmentation in local labor-market shapes the local effects of export shocks in a developing country. We first develop a theoretical framework that embeds trade and gender-segmented labor markets to show that foreign demand shocks may either increase or decrease the female-to-male employment ratio. The key theoretical result shows formally that the effects of trade on gender-segmented labor markets depend crucially on (a) the sectors that face the foreign demand shock; and (b) the domestic relevance of the foreign countries in which the demand shocks originate from. If the foreign demand shock from a relevant market happens in a female-intensive (male-intensive) sector, the model predicts that the female-to-male employment ratio should increase (decrease).

We then use plausibly exogenous variation exposure of Tunisian local labor markets to foreign demand shocks and show that the empirical results are consistent with the theoretical prediction. In Tunisia, a country with a high degree of gender segmentation in labor markets, foreign-demand shocks have been relatively larger in male-intensive sectors. This induced a decrease in the female-to-male employment ratio. Since male-intensive sectors had relatively more favorable foreign demand shocks, the equilibrium response is that households likely substituted female for male labor supply. Estimates using data from 2006 to 2016 confirm the theoretical mechanism postulated in this paper.

One important policy implication of our study is that less gender segmentation in labor markets will dampen the effect of foreign demand shocks on gender inequality. Taking foreign demand shocks as exogenous, the policymaker can only induce changes in the institutions that generate domestic gender segmentation in labor markets. If every sector has no gender segmentation, however, any foreign shocks would be distributionally neutral across genders – that is, males and females would be equally affected and allow more women to benefit from exports.

Thus, as countries develop their trade policies, they might want to consider policies that reduce gender segmentation in labor markets. Policies that promote gender equity have the immediate benefit of more gender equity in the present – but they can also have the unintended benefit of inducing more equitable effects in the future, whenever economic shifters affect the local economy.

\newpage

\printbibliography

\newpage 

\newpage
\appendix

\section{Details on data construction} \label{appendix: data}

\paragraph{Labor Market Survey} Labor market indicators come from the Tunisian Government’s National Survey on Population and Employment (\textit{Enquête Nationale sur la Population et l'Emploi} \textemdash ENPE). The survey is a repeated cross-section of households that has as its main goal producing statistics regarding the social, educational, and economic characteristics of the population that is either working, unemployed, or out of the labor force. Among other information, the ENPE reports individuals' employment status by industry (if any), gender, and region for seven years between 2006 and 2016. ENPE provides individual-level weights that aggregate to representative statistics of the Tunisian population. To construct the labor market indicators used in the empirical part of this paper, we excluded individuals under 15 years of age and aggregated the data at the region-year level using the provided weights.

\paragraph{Trade in Services} Data for trade in services the \href{https://www.wto.org/english/res_e/statis_e/trade_datasets_e.htm}{WTO-OECD BaTiS database}. The database uses the Extended Balance of Payments Services Classification (EBOPS) and inputing balance of payments data and harmonizes data to correct for divergences between import and exports of services. The data covers twelve sectors but for Tunisia only nine of them are relevant (represent $>99\%$ of trade in services). We created a concordance between EBOPS codes and ENPE codes.

\begin{table}[htp!]
\centering
\begin{tabular}{lll}
BaTiS Code & BaTiS id & Code ENPE  \\ \hline
SC         & 205      & 76        \\
SD         & 236      & 79        \\
SE         & 249      & 69        \\
SF         & 253      & 82        \\
SG         & 260      & 82        \\
SH         & 262      & 76        \\
SJ         & 268      & 85        \\
SK         & 287      & 89        \\
SL         & 291      & 93        \\ \hline
\end{tabular}
\end{table}

\paragraph{Trade in Merchandise} Data for trade in merchandise comes from \href{http://comtrade.un.org}{UNCOMTRADE}. In particular, we bulk downloaded the dollar value of Tunisian U.S. dollar exports at the Harmonized System 6-digit level (HS6) and then mapped HS codes into International Standard of Industrial Classification (ISIC) industry codes using the HS-ISIC \href{https://wits.worldbank.org/product_concordance.html}{concordances} publicized by the World Integrated Trade Solution (WITS). We then mapped ISIC codes into ENPE sectors. ENPE's sector classification is a national adaptation of the EU's Nomenclature of Economic Activities (NACE) codes. We mapped ISIC codes into NACE and ENPE codes using the following concordance matrix.

\begin{table}[htp!]
\centering

\begin{tabular}{lll}
ISIC Division Code & Chapter NACE & Code ENPE \\ \hline
1-3                & A            & 0         \\
5-9                & B            & 65        \\
10-12              & CA           & 10        \\
13-15              & CB           & 50        \\
16-19              & C            & 60        \\
20                 & CE           & 40        \\
21-22              & C            & 60        \\
23                 & CG           & 20        \\
24-25              & C            & 60        \\
26                 & CI           & 30        \\
27                 & CJ           & 30        \\
28                 & CK           & 30        \\
29-33              & C            & 60        \\
35                 & D            & 67        \\
36-39              & E            & 68        \\
41-43              & F            & 69        \\
45-47              & G            & 72        \\
49-53              & H            & 76        \\
55-56              & I            & 79        \\
58-63              & J            & 76        \\
64-66              & K            & 82        \\
68                 & L            & 85        \\
72-75              & M            & 85        \\
77-82              & N            & 85        \\
84                 & O            & 93        \\
85                 & P            & 93        \\
86-88              & Q            & 89        \\
90-93              & R            & 89        \\
94-96              & S            & 89        \\
97-98              & T            & 99        \\
99                 & U            & 98       \\
\hline
\end{tabular}
\end{table}

\newpage

\section{Summary Statistics}\label{appendix: summary}

\begin{table}[htp]
\centering
\begin{tabular}{lll}
\hline
\textbf{District}                   & \textbf{2008} & \textbf{2016} \\ \hline
Tunis                               & 5.52          & 4.96          \\
Ariana                              & 4.57          & 4.69          \\
Ben Arous                           & 4.73          & 4.52          \\
Manouba                             & 4.11          & 4.28          \\
Nabeul                              & 5.95          & 5.36          \\
Zaghouan                            & 3.46          & 3.79          \\
Bizerte                             & 5.09          & 4.37          \\
Beja                                & 3.95          & 3.9           \\
Jendouba                            & 3.48          & 3.15          \\
Le Kef                              & 3.78          & 4.1           \\
Siliana                             & 3.71          & 3.55          \\
Sousse                              & 4.5           & 4.6           \\
Monastir                            & 4.64          & 5.03          \\
Mahdia                              & 4.42          & 4.28          \\
Sfax                                & 5.91          & 7             \\
Kairouan                            & 4.42          & 4.37          \\
Kasserine                           & 3.24          & 3.6           \\
Sidi Bouzide                        & 4.59          & 3.39          \\
Gabes                               & 3.31          & 3.71          \\
Mednine                             & 4.11          & 4.12          \\
Tataouine                           & 3.7           & 3.06          \\
Gafsa                               & 3.04          & 3.6           \\
Tozeur                              & 2.58          & 3.25          \\
Kebili                              & 3.21          & 3.3           \\ \hline
\multicolumn{1}{l}{\textbf{Total}} & \textbf{100}  & \textbf{100}  \\ \hline
\end{tabular}

\caption{Distribution of Workers across regions. Pearson correlation across regions between periods: $\rho = .83$}
\end{table}

\begin{table}[htp]
\centering
\begin{tabular}{lll}
\hline
\textbf{Industry Code (NACE Rev. 3)}         & \textbf{2006} & \textbf{2016} \\ \hline
Agriculture, hunting, and fisheries          & 24.01         & 17.87         \\
Manufacture of food products                 & 1.92          & 2.25          \\
Manufacture of construction products      & 1.2           & 1.2           \\
Manufacture of Mechanical and Electrical Equipment      & 2.74          & 3.97          \\
Manufacture of chemicals                     & 0.71          & 0.84          \\
Manufacture of textiles and textile products & 7.73          & 5.5           \\
Other manufacture                            & 2.48          & 2.1           \\
Mining                                       & 0.24          & 0.24          \\
Extraction of oil and gas                    & 0.17          & 0.32          \\
Production and distribution of electricity      & 0.36          & 0.34          \\
Production and distribution of water         & 0.26          & 0.27          \\
Public works and buildings                   & 12.61         & 14.56         \\
Retail trade                                 & 10.87         & 12.36         \\
Transportation                               & 4.37          & 5.19          \\
Telecommunications                           & 0.9           & 0             \\
Hotels and Restaurants                       & 3.54          & 3.04          \\
Banking and Insurance                        & 0.68          & 0.68          \\
Real estate, repairs, and entrepreneurial activities      & 3.16          & 4.3           \\
Social and cultural services                 & 4.27          & 3.61          \\
Education and public administration          & 17.76         & 21.17         \\
Extraterritorial activity                    & 0.03          & 0.19          \\ \hline
\multicolumn{1}{l}{\textbf{Total}}          & \textbf{100}  & \textbf{100}  \\ \hline 
\end{tabular}

\caption{Distribution of Workers across regions. Pearson correlation across industries between periods: $\rho = .95$}
\label{table:industries}
\end{table}

\begin{table}[htp]
\centering

\begin{tabular}{lll}\hline
\textbf{Female Labor Share} & \textbf{2006} & \textbf{2016} \\ \hline
\textbf{Grand Tunis}        & 27.6\%        & 33.3\%        \\
\textbf{Northeast}          & 28.2\%        & 30.0\%        \\
\textbf{Northwest}          & 28.6\%        & 23.8\%        \\
\textbf{Mid-East}           & 30.0\%        & 30.0\%        \\
\textbf{Mid-West}           & 23.9\%        & 23.2\%        \\
\textbf{Southeast}          & 17.5\%        & 22.4\%        \\
\textbf{Southwest}          & 22.2\%        & 24.6\%       \\ \hline
\end{tabular}
\caption{Distribution of female labor shares across larger regions of Tunisia. Pearson correlation across industries between periods: $\rho = .66$}
\label{table:female-share-grand-regions}
\end{table}

\begin{table}[htp]
\resizebox{\textwidth}{!}{

\begin{tabular}{lll} \hline
\textbf{Industry} & \textbf{\begin{tabular}[c]{@{}l@{}}Female employment\\ share in 2006\end{tabular}} & \textbf{\begin{tabular}[c]{@{}l@{}}Export Growth \\ Contribution\\ Between 2006-16\end{tabular}} \\ \hline
\textbf{Manufacture of textiles and textile products}          & 74.9\% & -3.0\% \\
\textbf{Social and cultural services}                          & 45.0\% & 0.0\%  \\
\textbf{Banking and Insurance}                                 & 35.1\% & 0.0\%  \\
\textbf{Education and public administration}                   & 32.1\% & 0.1\%  \\
\textbf{Manufacture of Mechanical and Electrical Products}     & 28.9\% & 19.1\% \\
\textbf{Agriculture, hunting, and fisheries}                   & 27.9\% & 1.9\%  \\
\textbf{Manufacture of chemicals}                              & 22.3\% & 0.7\%  \\
\textbf{Manufacture of food products}                          & 19.7\% & 4.2\%  \\
\textbf{Real estate, repairs, and entrepreneurial services}    & 15.0\% & 0.2\%  \\
\textbf{Hotels and Restaurants}                                & 11.4\% & 8.8\%  \\
\textbf{Other manufacture}                                     & 11.4\% & 12.8\% \\
\textbf{Mining}                                                & 9.4\%  & -3.9\% \\
\textbf{Manufacture of construction products, clay, and glass} & 8.4\%  & 0.0\%  \\
\textbf{Production and distribution of water}                  & 7.3\%  & -1.4\% \\
\textbf{Transportation}                                        & 4.6\%  & 4.5\%  \\
\textbf{Public works and buildings}                            & 1.1\%  & 1.3\% \\ \hline
\end{tabular}

}
\caption{Relationship between initial female employment share and exports contribution growth}
\label{table:growth-contribution}

\end{table}

\begin{table}[]
\centering
\begin{tabular}{llllllll} \hline
Merchandise Exports & \textbf{2006} & \textbf{2008} & \textbf{2010} & \textbf{2011} & \textbf{2013} & \textbf{2015} & \textbf{2016} \\\hline
\textbf{France}         & 32\%  & 29\%  & 29\%  & 31\%  & 26\%  & 29\%  & 32\%  \\
\textbf{Italy}          & 22\%  & 21\%  & 20\%  & 22\%  & 18\%  & 18\%  & 17\%  \\
\textbf{Germany}        & 8\%   & 7\%   & 8\%   & 9\%   & 9\%   & 10\%  & 11\%  \\
\textbf{Spain}          & 6\%   & 5\%   & 4\%   & 4\%   & 5\%   & 5\%   & 3\%   \\
\textbf{Libya}          & 5\%   & 5\%   & 4\%   & 4\%   & 5\%   & 4\%   & 3\%   \\
\textbf{United Kingdom} & 3\%   & 5\%   & 5\%   & 3\%   & 4\%   & 3\%   & 2\%   \\
\textbf{Algeria}        & 2\%   & 2\%   & 3\%   & 3\%   & 3\%   & 4\%   & 5\%   \\
\textbf{ROW}            & 19\%  & 25\%  & 23\%  & 21\%  & 26\%  & 21\%  & 22\%  \\ \hline
\textbf{Total}          & 100\% & 100\% & 100\% & 100\% & 100\% & 100\% & 100\% \\ \hline
\end{tabular}

\caption{\textbf{Destination of Merchandise Exports from Tunisia, by Country.} Authors' calculations with UNCOMTRADE data. Calculated as exports reported by Tunisia having as destination all countries in the world.}
\label{table:destination-exports}
\end{table}

\newpage

\section{Proofs}

\paragraph{Proposition 1}

\begin{proof}
    Calculate $\frac{\partial L_{o,f} / L_{o,m}}{\partial e_d}$ for an arbitrary $d$. By the envelope theorem, since $e_d$ is a parameter, it locally only impacts consumption through a direct effect and any indirect effect due to reoptimization is zero. Therefore:
    
    \begin{equation*}
        \frac{\partial L_{o,f} / L_{o,m} }{\partial e_d} = \left\{ \pi_{od,f} (1-\alpha_d) \left[ \frac{\Xi_{o}}{1+\Xi_o} - \beta_{o,f}   \right] + \pi_{od,m} \alpha_d \left[ \frac{\Xi_{o}}{1+\Xi_o} - \beta_{o,m}  \right] \right\} \cdot \frac{P_o}{{\Xi_o^2}} \cdot \left(\frac{w_{o,f}}{w_{o,m}}\right)^{-1}
    \end{equation*}
    
For necessity, first note that, for a large enough $\alpha_d$ and sufficiently large enough $\pi_{od,m}$:

\begin{equation*}
   \left| \pi_{od,m} \alpha_d \left[ \frac{\Xi_{o}}{1+\Xi_o} - \beta_{o,m}  \right] \right| >  \left| \pi_{od,f} (1-\alpha_d) \left[ \frac{\Xi_{o}}{1+\Xi_o} - \beta_{o,f}   \right] \right|
\end{equation*}

\noindent which implies that $\frac{\partial L_{o,f}/ L_{o,m} }{\partial e_d} < 0$, because $\beta_{o,f} < \frac{\Xi_{o}}{1+\Xi_o} <  \beta_{o,m}$ and  $\frac{P_o}{{\Xi_o^2}} \cdot \left(\frac{w_{o,f}}{w_{o,m}}\right)^{-1} > 0$. Therefore, relative labor demand will decrease.

For sufficiency, assume that $\frac{\partial L_{o,f}/ L_{o,m} }{\partial e_d} < 0$. Then, since $\left(\frac{w_{o,f}}{w_{o,m}}\right)$, $P_o$, and $\Xi_o^2$ are positive numbers, this implies:

\begin{equation*}
    \underbrace{\pi_{od,f} (1-\alpha_d) \left[ \frac{\Xi_{o}}{1+\Xi_o} - \beta_{o,f}   \right]}_{>0} + \underbrace{\pi_{od,m} \alpha_d \left[ \frac{\Xi_{o}}{1+\Xi_o} - \beta_{o,m}  \right]}_{<0} < 0
\end{equation*}

which requires a sufficiently large $\pi_{od,m}$ and a sufficiently large $\alpha_d$. 
\end{proof}

\end{document}